\documentclass[onecolumn,authoryear]{els-mrw} 
\usepackage{caption} 
\usepackage{amsmath,amssymb,amsfonts,amsthm,makeidx,graphicx}
\usepackage{txfonts}
\usepackage{helvet}
\usepackage{slashed}
\definecolor{darkgreen}{rgb}{0.0, 0.5, 0.0}
\definecolor{orange}{rgb}{1.0, 0.5, 0.0}
\newcommand{\bea}{\begin{eqnarray}}
\newcommand{\eea}
{\end{eqnarray}}
\usepackage{cancel}
\newcommand{\CPv}{\ensuremath{\cancel{\mathrm{CP}}}}

\begin{document}

\chapter{The theory of electric dipole moments: the view from below}\label{chap1}

\author[1,2]{Jordy de Vries}%

\address[1]{\orgname{University of Amsterdam}, \orgdiv{Institute of Theoretical Physics}, \orgaddress{ Science Park 904, 1098 XH Amsterdam, The Netherlands}}
\address[2]{\orgname{Nikhef}, \orgdiv{Theory department}, \orgaddress{Science Park 105, 1098 XG, Amsterdam, The Netherlands}}

%\articletag{Chapter Article tagline: update of previous edition,, reprint..}

\maketitle

%\begin{glossary}[Glossary]
%\term{Europe} the model is a coherent view of capital markets data that allows users to interact with the content in a consistent manner.

%\term{Primates} regardless of the source. Essentially, of sources. Properly deployed.

%\end{glossary}

%\begin{glossary}[Nomenclature]
%\begin{tabular}{@{}lp{34pc}@{}}
%AF &Assessment Factor\\
%ECHA &European Chemical Agency\\
%EPM &Equilibrium Partitioning Method Equilibrium Partitioning Method Equilibrium Partitioning Method Equilibrium\hfill\break Partitioning Method\\
%ERA &Ecological Risk Assessment\\
%HC &Hazardous Concentration\\
%\end{tabular}
%\end{glossary}

\begin{abstract}[Abstract]
Permanent electric dipole moments (EDMs) of nucleons, nuclei, atoms, and molecules 
are among the most sensitive probes of CP violation beyond the Standard Model and 
are intimately connected to the strong CP problem and the origin of the 
matter-antimatter asymmetry of the universe. This review presents the theory of EDMs 
from the bottom up, tracing the chain of connections that links CP-violating 
interactions at level of elementary particles to observable EDMs across a wide range of 
systems. Starting from a general CP-odd effective Lagrangian at the quark-gluon level comprising the QCD $\bar\theta$ term, quark EDMs and chromo-EDMs, the 
Weinberg operator, and CP-odd four-fermion interactions, I show how chiral 
perturbation theory organizes the nonperturbative QCD dynamics into a small set of 
hadronic low-energy constants, whose relative sizes are 
determined by the chiral representation of the underlying source. These hadronic 
interactions feed into  calculations of nuclear EDMs and Schiff moments, 
which in turn enter atomic and molecular structure calculations that connect to 
experimentally accessible observables in diamagnetic and paramagnetic systems. 
 Special attention is given to 
the recently identified sensitivity of paramagnetic systems to hadronic CP violation, 
which opens a new and relatively unexplored window on the quark-gluon sector. The 
complementarity of the full EDM portfolio including the neutron, light nuclei, 
atoms, and molecules, and the role of theory in disentangling the underlying source of CP violation 
is discussed throughout.
\end{abstract}

\section{Introduction}\label{chap1:sec1}

Permanent electric dipole moments (EDMs) of elementary particles, nucleons, nuclei, atoms, and molecules are among the most sensitive probes of CP violation available. An EDM is the analogue of a magnetic dipole moment, but instead of the coupling between spin and a magnetic field it couples to an electric field. Its existence requires both parity (P) and time-reversal (T) violation, and by the CPT theorem, CP violation. Due to its flavor structure, the single known source of CP violation in the quark-mixing matrix, predicts EDMs that are extraordinarily small, many orders of magnitude below current experimental sensitivity. Therefore, any EDM observed in the foreseeable future would be unambiguous evidence for new sources of CP violation. This is not idle hope as additional CP violation is expected to exist to explain the asymmetry between matter and antimatter in our universe. 

\begin{figure}[t!]
    \centering
    \includegraphics[width=0.8\textwidth]{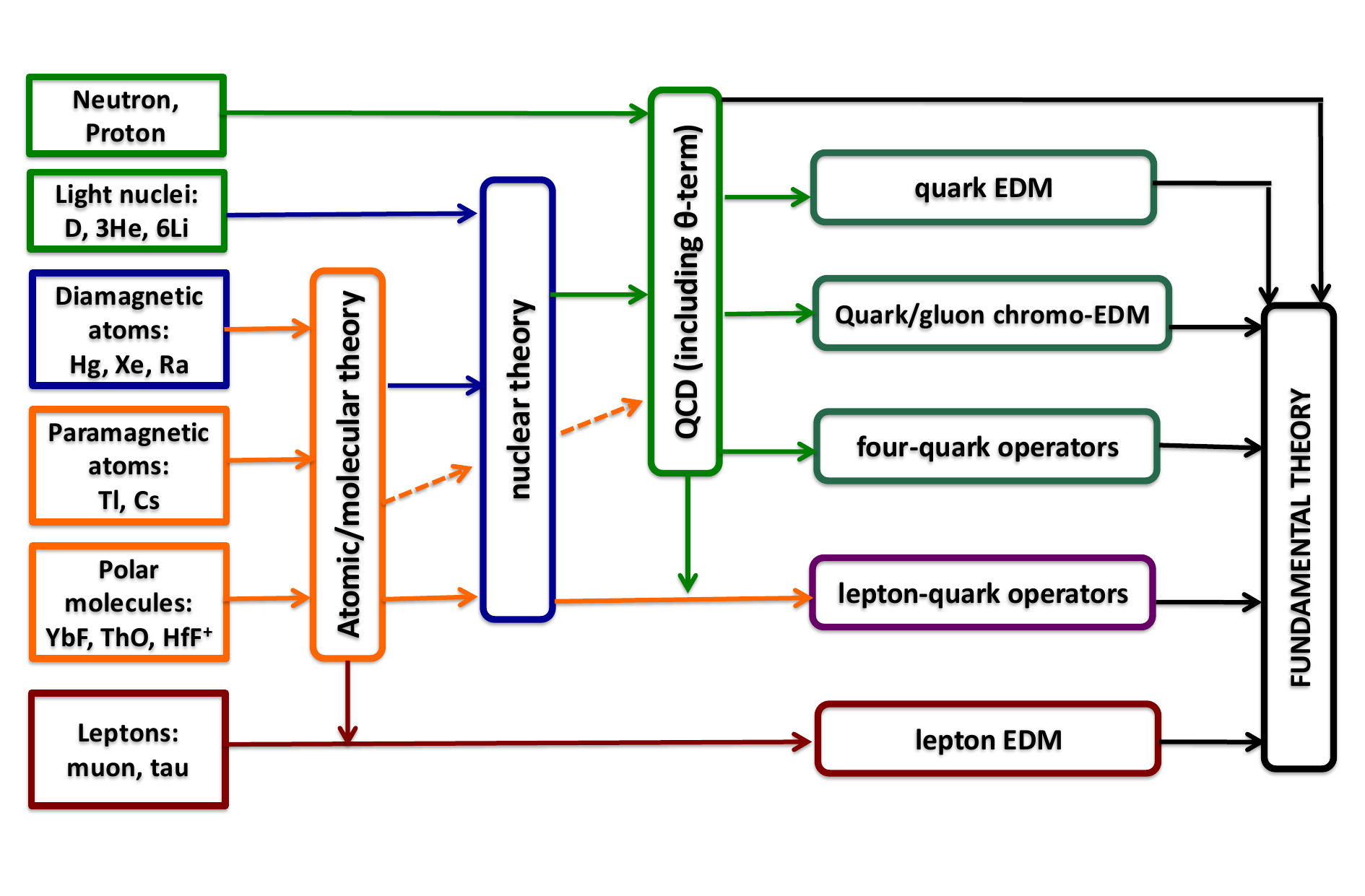}
    \caption{The EDM `metro map' showing how to connect the EDM measurements (left) to the fundamental theory of CP violation (right) by going through molecular, atomic, nuclear, and hadronic theory. Effective operators such as  quark EDMs and chromo-EDMs are used as intermediate steps. The dashed orange arrows show connections that are recently identified and reviewed in Sect.~\ref{sec:paratodia}.}
    \label{fig:landscape}
\end{figure}

The experimental landscape has advanced significantly in recent years. The best limit on the electron EDM, $|d_e| < 4.1 \times 10^{-30}$ e cm, comes from precision spectroscopy of the polar molecule HfF$^+$ (\cite{Roussy:2022cmp}), with competitive results from ThO (\cite{ACME:2018yjb}). The neutron EDM is bounded at $|d_n| < 1.8 \times 10^{-26}$ e,cm~(\cite{Abel:2020pzs}), and the diamagnetic atom ${}^{199}$Hg , $d_{{}^{199}\mathrm{Hg}} < 7.4 \cdot 10^{-30}\,e\,\mathrm{cm}$,  provides the tightest constraint on hadronic CP violation in heavy nuclei (\cite{PhysRevLett.116.161601}). Next-generation experiments, including new molecular EDM searches, improved neutron EDM measurements, storage ring experiments targeting proton and nuclear EDMs, and searches in radioactive species such as ${}^{225}$Ra and RaF, aim to improve the sensitivity over the coming decade (\cite{Alarcon:2022ero}). While the experimental limits in these classes differ by several orders of magnitude, their sensitivity to underlying quark, gluon, and semi-leptonic operators is often comparable, owing to a web of enhancement and suppression factors associated with finite-size effects, violations of Schiff shielding, nuclear collectivity, and molecular polarization. Navigating this web, and ultimately interpreting a future EDM signal in terms of a specific source of CP violation, requires a quantitative theoretical framework connecting fundamental CP violation to hadronic, atomic, and molecular observables, see Fig.~\ref{fig:landscape}. The aim of this review is to describe that framework.

EDMs are intimately connected to the strong CP problem, one of the fine-tuning puzzles in the Standard Model (SM). The QCD Lagrangian admits a CP-violating term proportional to $\bar\theta$, the sum of the vacuum angle and the argument of the quark mass determinant. Experimental bounds on the neutron EDM  constrain $|\bar\theta| \lesssim 10^{-10}$, yet there is no symmetry reason within the SM for $\bar\theta$ to be this small. The most elegant solution is the Peccei-Quinn mechanism (\cite{Peccei:1977hh}), which promotes $\bar\theta$ to a dynamical field relaxed to zero by a new pseudo-Nambu-Goldstone boson, the axion. EDM experiments are sensitive to $\bar\theta$ directly, and as discussed in this review, the pattern of EDMs across different systems might reveal whether the strong CP problem is solved by an axion mechanism or in some other way. 

The network of connections between microscopic CP‑odd sources and EDM observables is naturally described within an effective field theory (EFT) framework. At and above the electroweak scale, one writes a general CP‑odd EFT in terms of SM fields, organized in operators of increasing dimension, and matches this onto a basis of quark, gluon, and lepton operators at a hadronic scale of order 1–2 GeV.  A recent review (\cite{Pospelov:2025vzj}) provides a broad overview of EDMs as probes of physics beyond the SM (BSM), emphasizing this operator basis at and above the electroweak scale and the connection to a wide range of ultraviolet (UV) models such as supersymmetry, extended Higgs sectors, neutrino portals, and dark sectors. Another recent review has focused on the role of EDMs in testing electroweak baryogenesis (\cite{vandeVis:2025efm}). 

The present review is intended as a complement to these particle physics 
and cosmological perspectives. While a bird's eye view sees the big picture and surveys the broad 
landscape of BSM scenarios, the worm's eye view from 
the mud of hadronic, nuclear, atomic, and molecular physics can reveal 
structures that are invisible from the top. Rather than surveying UV model space, I therefore concentrate on the hadronic, nuclear, and atomic/molecular layers of the EFT hierarchy that connect CP‑violating sources at level of elementary particles to the EDMs of nucleons, nuclei, atoms, and molecules. The starting point is an effective CP‑odd Lagrangian formulated at a renormalization scale around a few GeV, containing the QCD $\theta$-term, quark and lepton EDMs, quark chromo‑EDMs, the three‑gluon (Weinberg) operator, and CP-violating four‑fermion operators. Below this scale, the appropriate degrees of freedom are hadrons rather than quarks and gluons, and nonperturbative QCD is encoded in the couplings of a chiral EFT involving pions, nucleons, and heavier baryons. The hadronic interactions in turn feed into nuclear many‑body calculations, which determine the EDMs and Schiff moments of nucleons, light and heavy nuclei, and ultimately into atomic and molecular structure calculations that connect to atomic and molecular EDMs. Throughout, I emphasize the progress made  based on EFTs and modern nuclear and atomic/molecular many‑body methods, which aims to organize these steps in a controlled expansion and to provide a consistent language for comparing different systems and sources. I put extra emphasis on recent work connecting hadronic CP violation to paramagnetic systems as this is still rather unexplored territory.

\section{CP-violating effective  interactions at low energy}
\label{sec:CPVLag}

The interpretation of EDM searches in terms of fundamental sources of CP violation involves a multitude of scales ranging from molecular (eV) to BSM physics (TeV or beyond). Above the electroweak scale, the appropriate language is the Standard Model Effective Field Theory (SMEFT) (\cite{Buchmuller:1985jz}, \cite{Grzadkowski:2010es}), in which BSM physics at a scale $\Lambda \gg v$ is integrated out and encoded in a tower of $\mathrm{SU}(3)_c \times \mathrm{SU}(2)_L \times \mathrm{U}(1)_Y$ gauge-invariant operators built from SM fields. CP-odd effects first appear at dimension six, so the leading BSM contributions to CP violation carry a suppression $v^2/\Lambda^2$ relative to the SM, where $v \simeq 246$ GeV is the Higgs vacuum expectation value. The requirement of full gauge invariance is a powerful constraint on the operator basis and strongly constrains the possible CP-odd interactions. 

The effective SMEFT operators involve the full SM field content. Below the electroweak scale, heavy degrees of freedom such as the Higgs, top quarks, and electroweak gauge bosons, are integrated out, leaving a tower of local CP-odd operators built from light SM fields. Renormalization-group (RG) evolution from the electroweak scale down to $\mu \sim 1$-$2$~GeV, where quarks and gluons are matched onto hadronic degrees of freedom, reshuffles these operators and generates a correlated ensemble of Wilson coefficients, see Sect.~\ref{sec:RGE} for a discussion. At the hadronic matching scale, around 2 GeV, the CP-violating Lagrangian takes the form
\begin{equation}
\mathcal{L}_{\rm CPV} = \mathcal{L}_{\bar\theta} + \mathcal{L}_{d_q} + \mathcal{L}_{\bar d_q} + \mathcal{L}_{W} + \mathcal{L}_{4q} + \mathcal{L}_{d_e}+  \mathcal{L}_{\rm sl} + \ldots \,,
\label{eq:LCPVfull}
\end{equation}
where the dots denote higher-dimensional operators. In this section I introduce each term in turn, with enough detail to map to a theory of hadronic CP violation in Sect.~\ref{sec:chiralEFT}.

\subsection{The QCD $\bar\theta$ term}

The only CP-odd interaction of dimension four in QCD is the topological term (\cite{tHooft:1976rip})
\begin{equation}
\mathcal{L}_{\bar\theta} = -\bar\theta \, \frac{g_s^2}{32\pi^2} \, G^a_{\mu\nu} \tilde G^{a\,\mu\nu} \,,
\label{eq:Ltheta}
\end{equation}
where $G^a_{\mu\nu}$ is the gluon field-strength tensor, $\tilde G^{a\,\mu\nu} = \frac{1}{2}\epsilon^{\mu\nu\alpha\beta} G^a_{\alpha\beta}$ its dual, and $\bar\theta = \theta_{\rm QCD} + \arg\det M_q$ is the physical combination of the QCD vacuum angle and the phase of the quark-mass matrix $M_q$. The operator $G\tilde G$ is a total derivative and has no perturbative effects, but leads to non-perturbative hadronic matrix elements contributing to, for example, the neutron EDM.

The strong CP problem is the question of why $|\bar\theta| \lesssim 10^{-10}$, as required by the neutron EDM bound (\cite{Abel:2020pzs}). While several mechanisms have been proposed to explain why $\bar \theta$ is so small,  $\bar\theta$ remains a legitimate free parameter of the low-energy theory.
Even when a Peccei-Quinn (\cite{Peccei:1977hh}) mechanism is present, an effective $\bar\theta$ is induced by other CP-odd sources present in the theory once the axion field takes its minimum value (see for example \cite{Pospelov:2000bw} and \cite{Dekens:2022gha}). The hadronic consequences of the $\bar\theta$ term must be understood in any case.

The gluonic term in $\mathcal{L}_{\bar\theta}$ can be traded, via an axial $U(1)$ rotation of the quark fields, for a complex contribution to the quark mass matrix
\begin{equation}
\mathcal{L}_{\bar\theta} \,\longrightarrow\, \bar\theta\, m_* \, \bar q i\gamma_5 q \,,
\label{eq:thetamass}
\end{equation}
where $q = (u\,d)^T$ and $m_* = m_u m_d/(m_u + m_d)$ is the reduced quark mass. This form makes the chiral transformation properties explicit: the $\bar\theta$ term acts as an imaginary quark mass, vanishes in the chiral limit, and conserves isospin.

\subsection{Quark EDMs and chromo-EDMs}

The quark electric dipole moments (qEDMs) are dimension-five dipole operators
\begin{equation}
\mathcal{L}_{d_q} = -\frac{i}{2}\sum_{q=u,d,s} d_q \, \bar q \sigma^{\mu\nu}\gamma_5 q \, F_{\mu\nu} \,,
\label{eq:qEDM}
\end{equation}
where $F_{\mu\nu}$ is the electromagnetic field strength. The qEDM is a chirality-odd operator that transforms in the same chiral representation as the quark mass matrix, but it is tied to the electromagnetic current. As a consequence, at hadronic scales it primarily induces operators with explicit photons, such as nucleon EDMs, while purely hadronic interactions are suppressed by $\alpha_{\rm em}$. This makes the qEDM phenomenologically distinct from the $\bar\theta$ term and quark chromo-EDMs (qCEDMs) despite sharing the same chiral representation.

The quark chromo-EDMs are the gluonic analogs of the qEDMs
\begin{equation}
\mathcal{L}_{\bar d_q} = -\frac{i}{2}\sum_{q=u,d,s} \bar d_q \, \bar q \sigma^{\mu\nu}\gamma_5 t^a q \, g_s G^a_{\mu\nu} \,,
\label{eq:qCEDM}
\end{equation}
with $t^a$ the $\mathrm{SU}(3)_c$ generators. The qCEDMs share the chiral transformation properties of the $\bar\theta$ term and quark masses, but couple to gluons rather than photons. Their low-energy phenomenology is therefore much closer to that of the $\bar \theta$ term. An important difference is that the qCEDMs can break isospin if $\bar d_u \neq \bar d_d$.

It is important to stress that, although Eqs.~\eqref{eq:qEDM} and \eqref{eq:qCEDM}
are dimension-five operators in the low-energy QCD+QED theory, they originate
from \emph{dimension-six} operators in the Standard Model EFT (SMEFT). The
reason is that above the electroweak scale the theory is invariant under
$\text{SU}(2)_L \times \text{U}(1)_Y$, so chirality-flipping dipole operators
must involve the Higgs field.
 The Wilson coefficients $d_q$ and $\bar d_q$ are therefore effectively suppressed as $v/\Lambda^2$ where $\Lambda$ is the BSM scale. In many explicit BSM scenarios an additional small Yukawa coupling  appears effectively replacing $v\rightarrow m_q$, the light quark mass, but this is not true for all scenarios (\cite{Dekens:2018bci}). The CP-odd quark dipoles are induced in many BSM models. Perhaps the most famous example is supersymmetry where they can be induced at one loop already through the exchange of virtual supersymmetric particles (\cite{Ellis:1982tk,Buchmuller:1982ye}).

\subsection{The Weinberg three-gluon operator}

The CP-odd three-gluon (Weinberg) operator is the unique dimension-six purely gluonic CP-violating operator (\cite{Weinberg:1989dx})
\begin{equation}
\mathcal{L}_W = \frac{C_W}{6}\, f^{abc} \, \epsilon^{\alpha \beta \mu \nu} G^a_{\alpha\beta}  G^{b}_{\mu \rho} \, G^{c\,\rho}_\nu \,,
\label{eq:Weinberg}
\end{equation}
with $f^{abc}$ the $\mathrm{SU}(3)_c$ structure constants. The Weinberg operator can be interpreted as the chromo-electric dipole moment of the gluon (\cite{Braaten:1990zt}). Unlike the $\bar\theta$ term,  $\mathcal{L}_W$ generates CP-odd vertices at the perturbative level and does not vanish in the chiral limit. Because it involves only gluon fields and no quark bilinears, the Weinberg operator conserves chiral symmetry making its low-energy phenomenology rather different from the $\bar \theta$ term and q(C)EDMs. The Weinberg operator is induced after integrating out heavy quarks with a qCEDM (\cite{Braaten:1990zt}) and thus appear often together with q(C)EDMs. It can also be generated in multi-Higgs (\cite{Weinberg:1989dx}) and leptoquark models (\cite{Abe:2017sam}).

\subsection{Four-quark operators}

At dimension six, flavor-diagonal four-quark operators provide two physically distinct classes of CP-odd sources. The first class consists of chiral-invariant operators of the form
\begin{equation}
\mathcal{L}^{(q\text{-inv})}_{4q} \sim C^{(1)}_{4q} \left[(\bar q q)(\bar q i\gamma_5 q)- (\bar q \vec \tau q)\cdot (\bar q \vec \tau i\gamma_5 q)\right] + C^{(8)}_{4q} \left[(\bar q t^a q)(\bar q i\gamma_5 t^a q)-(\bar q \vec \tau\,t^a q)\cdot (\bar q \vec \tau\i\gamma_5 t^a q)\right] \,,
\label{eq:4qinv}
\end{equation}
where $q=\{u\,d\}^T$. More operators appear once strange quarks are included. 
The operators in Eq.~\eqref{eq:4qinv} share a common low-energy phenomenology with the Weinberg operator. They are generated directly in the SMEFT from four-quark operators and are for example induced in leptoquark models (\cite{Dekens:2018bci}).

The second and phenomenologically more distinctive class is the four-quark left-right (FQLR) operator (\cite{Ng:2011ui})
\begin{equation}
\mathcal{L}^{(\rm LR)}_{4q} = C_{\rm LR}\, i \left(\bar u_R \gamma^\mu d_R \, \bar d_L \gamma_\mu u_L - \bar d_R \gamma^\mu u_R \, \bar u_L \gamma_\mu d_L \right) + \ldots
\label{eq:FQLR}
\end{equation}
where the dots denote similar operators with different color structure and/or strange quarks. The FQLR transforms as $(3_L, 3_R)$ under $\mathrm{SU}(2)_L \times \mathrm{SU}(2)_R$, which is a fundamentally different chiral representation than the $\bar\theta$ term or the qCEDMs, leading to a  qualitatively different low-energy phenomenology. The FQLR is not a direct SMEFT operator but is induced after electroweak symmetry breaking from the SMEFT operator $i(\bar H^\dagger D_\mu H)(\bar u_R \gamma^\mu d_R) + \mathrm{h.c.}$, which generates a coupling of the $W$-boson to right-handed quarks. Integrating out the $W$-boson then produces Eq.~\eqref{eq:FQLR}. This operator dominates EDM phenomenology in left-right symmetric models (\cite{Xu:2009nt}).

\subsection{Lepton EDMs and semi-leptonic operators}

The electron EDM (and analogously the muon and tau EDM)
\begin{equation}
\mathcal{L}_{d_e} = -\frac{i}{2} d_e \, \bar e \sigma^{\mu\nu} \gamma_5 e \, F_{\mu\nu} \,,
\label{eq:eEDM}
\end{equation}
is the primary target of paramagnetic EDM experiments and is a purely leptonic CP-odd source. It is generated in the SMEFT from the similar dimension-six operators as the qEDM, after inserting a Higgs vev. It appears in similar BSM scenarios as the qEDMs and qCEDMs. 

Beyond the electron EDM, atomic and molecular EDMs  are also sensitive to CP-odd interactions between electrons and quarks. At dimension six, three classes of dimension-six semi-leptonic operators are relevant for EDM phenomenology, distinguished by their Lorentz structure
\begin{equation}\label{Lsl}
\mathcal L_{\mathrm{sl}}= \bar e i \gamma^5 e\,(c_S^{(0)}\bar q q +  c_S^{(1)}\bar q \tau^3 q) + \bar e i \sigma^{\mu\nu}\gamma^5 e\,(c_T^{(0)}\bar q \sigma_{\mu\nu} q +  c_T^{(1)}\bar q \sigma_{\mu\nu} \tau^3 q) + \bar e  e\,(c_P^{(0)}\bar q i \gamma^5 q +  c_P^{(1)}\bar q i \gamma^5\tau^3 q)\,.
\end{equation}
 Similar to the chiral-invariant four-quark operators these operators appear directly in SMEFT from four-fermion operators and arise, for example, from tree-level leptoquark exchange (\cite{Dekens:2018bci,Fuyuto:2018scm}). Similar couplings to heavier quarks can be written which, after integrating out the heavy quarks, leads to operators of the form $\bar e e\,G^a_{\mu\nu}\tilde G^{\mu\nu,a}$ and $e \bar i \gamma^5 e\,G^a_{\mu\nu} G^{\mu\nu,a}$. 
The hadronization of the semi-leptonic operators, discussed in more detail below, is relatively straightforward and mainly leads to operators of the same for as in Eq.~\eqref{Lsl} but replacing $q\rightarrow N$, where $N=\{p\,n\}^T$ is the nucleon doublet.

\subsection{Renormalization-group running and mixing}
\label{sec:RGE}

In a realistic BSM scenario the operators in Eq.~\eqref{eq:LCPVfull} do not appear in isolation at the hadronic scale. They are generated at some high unknown energy scale and evolved down by QCD (and above the electroweak scale, electroweak) RG running. Two qualitative features of this evolution are important.
First, the Wilson coefficients receive multiplicative QCD corrections. For example, the Weinberg operator and the four-quark operators typically
receive sizeable multiplicative corrections when evolved from a multi-TeV
scale down to a few GeV (\cite{Braaten:1990zt}). In many cases these corrections are of order unity,
so they must be taken into account for quantitative work, but they do not
completely reshuffle the hierarchy among different sources. Calculations of these corrections have been extended to the two-loop level (\cite{Degrassi:2005zd}, \cite{deVries:2019nsu}, \cite{Naterop:2025cwg}).

Second, and often more relevant for phenomenology, operators mix under RG evolution. A four-quark operator at a high scale will induce quark EDMs and CEDMs at lower scales through quark-gluon loops. A heavy-quark CEDM generates a Weinberg operator below the heavy-quark threshold. Two-loop Barr-Zee diagrams connect CP-violating Higgs-fermion interactions to fermion EDMs and CEDMs (\cite{Barr:1990vd}, \cite{Brod:2023wsh}). The net result is that even a single CP-odd operator at the UV matching scale generates a correlated ensemble of operators at the hadronic scale. The full mixing and matching procedure is discussed in, for example, (\cite{Dekens:2013zca,Kley:2021yhn, Choi:2026tun}).
For this reason, EDM analyses at the hadronic, nuclear, atomic, and molecular level should ideally be formulated in terms of the full set of operators in Eq.~\eqref{eq:LCPVfull}. A global analysis of this type was performed in (\cite{ Gaul:2023hdd,Degenkolb:2024eve}) and illustrates the complementary of EDM experiments on different systems.

That being said, a useful  organizing principle for the various operators is based on their
transformation property under the approximate chiral symmetry $\mathrm{SU}(2)_L \times \mathrm{SU}(2)_R$ of QCD. As discussed in detail in Sect.~\ref{sec:chiralEFT}, these properties determine the form and relative sizes of CP-odd hadronic interactions which, in turn, determines the hierarchy of EDMs. 
 It is therefore useful to group the hadronic operators in Eq.~\eqref{eq:LCPVfull} into three distinct classes:
\begin{itemize}
\item \textbf{Mass-like sources} ($\bar\theta$, qEDMs, qCEDMs): these transform under $\mathrm{SU}(2)_L \times \mathrm{SU}(2)_R$ in the same representation as the quark mass matrix, $(2_L, \bar 2_R) \oplus (\bar 2_L, 2_R)$. They generate CP-odd pion-nucleon couplings already at leading order in the chiral expansion, and their hadronic consequences closely track those of the $\bar\theta$ term. The $\bar\theta$ term and the isoscalar qCEDM conserve isospin, the isovector qCEDM (with $\bar d_u \neq \bar d_d$) breaks it.
\item \textbf{Chiral singlets} (Weinberg operator, chirally invariant four-quark operators): these do not break chiral symmetry at the Lagrangian level. CP-odd pion-nucleon interactions therefore require additional derivatives or quark-mass insertions and are suppressed in the chiral expansion. Short-range nucleon-nucleon interactions play a comparatively larger role for these sources.
\item \textbf{Tensorial sources} (FQLR operator): these transform as $(3_L, 3_R)$ under $\mathrm{SU}(2)_L \times \mathrm{SU}(2)_R$. They generate a large isovector pion-nucleon coupling $\bar g_1$ at leading order and a three-pion vertex, leading to a phenomenology that is qualitatively distinct from the $\bar\theta$ term and from chiral singlets.
\end{itemize}
In addition there are the semi-leptonic operators that fall into their own class. This classification scheme will be used below to organize the discussion of nucleon, nuclear, and atomic/molecular EDMs.

\section{Chiral effective field theory for CP-odd sources}
\label{sec:chiralEFT}
Below $\mu \sim 1$ GeV the relevant degrees of freedom are no longer quarks and gluons but pions, nucleons, and heavier hadrons, and photons. The systematic framework for connecting the quark-gluon operators of Sect.~\ref{sec:CPVLag} to hadronic observables is chiral effective field theory ($\chi$EFT), the low-energy EFT of QCD (\cite{Weinberg:1978kz}, \cite{Gasser:1983yg}, \cite{Weinberg:1990rz}). Its great advantage is that the structure of the induced CP-violating hadronic interactions, \textit{i.e.} their form and relative sizes, is completely determined by symmetry even though the numerical values of the associated low-energy constants (LECs) require nonperturbative QCD input. $\chi$EFT also allows for a systematic calculation of higher-order loop corrections and can incorporate interactions with external currents such as electromagnetism and weak interactions. The construction of the CP-odd chiral Lagrangian in terms of the operators in Eq.~\eqref{eq:LCPVfull} has been worked out in (\cite{Mereghetti:2010tp}, \cite{deVries:2012ab}, \cite{Bsaisou:2014oka}). In this section I will explain the method in some detail for the $\bar\theta$ term, which serves as the canonical example, and then summarize the results for the other sources.
\subsection{Chiral symmetry and the spurion method}
\label{sec:spurion}
In the absence of electromagnetism and in the limit of vanishing light-quark masses, the QCD Lagrangian for two flavors is invariant under independent $\mathrm{SU}(2)$ rotations of the left- and right-handed quark doublet
\begin{equation}
q_L \to L\, q_L\,, \qquad q_R \to R\, q_R \, , \qquad L,R \in \mathrm{SU}(2)_{L,R} \, .
\end{equation}
This chiral symmetry $G = \mathrm{SU}(2)_L \times \mathrm{SU}(2)_R$ is spontaneously broken to the diagonal isospin subgroup $H = \mathrm{SU}(2)_V$, with the three pions as the associated pseudo-Goldstone bosons. They are collected in a unitary matrix
\begin{equation}
U(x) = \mathrm{exp}\left(\frac{i}{F_\pi} \vec \pi(x)\cdot \vec \tau\right)\,,
\end{equation}
in terms of the pion triplet $\vec \pi(x)$ and the pion decay constant $F_\pi \simeq 92.2$ MeV. The matrix transforms under chiral symmetry as $U \to L U R^\dagger$. 
The key tool for incorporating explicit symmetry breaking is the spurion method. Rather than treating symmetry-breaking terms as perturbations, one formally promotes their coefficients to spurion fields that transform under $G$ in such a way that  the full Lagrangian is chiral invariant. Physical results are recovered by setting the spurion to its actual (symmetry-breaking) value at the end. The power of the method is that it allows one to read off the complete set of allowed operators at any given order in the chiral expansion purely from group theory, without doing any explicit matching calculation.
A good example is the quark mass matrix $M = \mathrm{diag}(m_u, m_d)$. It breaks chiral symmetry because it couples left- and right-handed quarks. One promotes $M$ to a spurion transforming as $M \to L M R^\dagger$, so that the combination $\chi \equiv 2B\, M$ (with $B$ related to the quark condensate) can be inserted into the chiral Lagrangian to build invariants. At leading order in the mesonic sector this gives
\begin{equation}
\mathcal{L}_\pi = \frac{F_\pi^2}{4}\mathrm{Tr}[D_\mu U D^\mu U^\dagger] + \frac{F_\pi^2}{4}\mathrm{Tr}[\chi U^\dagger + \chi^\dagger U]\,,
\label{eq:LO_meson}
\end{equation}
where the first term is the pion kinetic energy in terms of covariant derivatives $D_\mu$ (in this work $D_\mu$ contains the couplings to photons) and the second term generates the pion mass $m_\pi^2 = B (m_u + m_d)$. Note that the pion mass is not predicted as it involves the non-perturbative LEC $B = \mathcal O(\Lambda_\chi)$, but $\chi$EFT does predict a linear dependence on the average quark mass. 

Nucleons are introduced as a doublet $N = (p\,n)^T$ transforming under the unbroken $\mathrm{SU}(2)_V$. Because the theory is based on an expansion in $Q/\Lambda_\chi$ where $Q\sim m_\pi$ is a low-energy scale and $\Lambda_\chi \sim 2\pi F_\pi \sim 1$ GeV, virtual nucleon fields inside loop diagrams can lead to factors of $m_N/\Lambda_\chi = \mathcal O(1)$ that break the power counting. This can be avoided by using the fact that nucleons (at least as far as EDM experiments are concerned) are non-relativistic and using a heavy-baryon (HB) description (\cite{Jenkins:1990jv}) where the nucleon mass is removed from the nucleon propagator. In such a HB framework it is the spin $S^\mu = (0 \,\vec \sigma/2)$ and velocity $v^\mu = (1\,\vec 0)$ in the rest frame that appears instead of gamma matrices. In the one-nucleon sector, for our purposes the most relevant Lagrangian in the HB formulation reads
\begin{equation}
\mathcal{L}_{\pi N} = \bar N \left(iv\cdot D + g_A S\cdot u\right) N + \bar N\left( c_1\,\mathrm{Tr}\,\chi_+  + c_5\,\hat \chi_+ \right)N \,.
\label{eq:LO_piN}
\end{equation}
In the first two terms, $u_\mu = i(u^\dagger \partial_\mu u - u\partial_\mu u^\dagger)$ with $u^2 = U$, and $g_A \simeq 1.27$ the axial coupling gives rise to the leading CP-conserving pion-nucleon interaction. In the last two terms $\chi_+ = u^\dagger \chi u^\dagger + u \chi^\dagger u$ and $\hat \chi_+ = \chi_+ - \mathrm{Tr}(\chi_+)/2$. The $c_1$ and $c_5$ terms appear at next-to-leading order, $c_{1,5}= \mathcal O(\Lambda_\chi^{-1})$, and describe quark mass contributions to the nucleon mass. In particular
\begin{eqnarray}
\Delta m_N &\equiv& \frac{(m_n + m_p)}{2} - m_N^0 = -4 B (m_u + m_d) c_1\,,\\
\delta m_n &\equiv&  = m_n -m_p =-4 B(m_d -m_u) c_5\,,
\end{eqnarray} 
where $m_N^0$ is the nucleon mass in the chiral limit and it is important to note that $\delta m_N$ is only the strong part of the nucleon mass splitting and does not include electromagnetic corrections.

\subsection{The $\bar \theta$ term as a worked out example}
\label{sec:theta_spurion}
Let us now apply the spurion method to the $\bar\theta$ term. As shown in Eq.~\eqref{eq:thetamass}, after an axial $U(1)$ rotation the $\bar\theta$ term becomes an imaginary contribution to the quark mass matrix. This means that the induced CP-odd hadronic interaction can be read from the usual chiral Lagrangian by simply replacing
\begin{equation}
\chi \rightarrow 2 B(M + i \bar \theta m_\star)\,.
\end{equation}
 Because the $\bar \theta$ term is proportional to the identity matrix in flavor space it leads to isospin-conserving CP-odd interactions. Let's consider the effects on the pion Lagrangian in Eq.~\eqref{eq:LO_meson}. The terms proportional to $\bar \theta$ are given by 
\begin{equation}
\frac{F_\pi^2}{4} (2i Bm_\star)\,\mathrm{Tr}[ U^\dagger -U] =0\,,
\end{equation}
which vanishes exactly. This implies that without insertions of further spurions, the $\bar \theta$ term does not induce CP-violating pionic interactions such as $\pi^0$ or $\vec \pi^{\,2}\,\pi^0$, a direct consequence of isospin conservation. 

In the one-nucleon sector we can include $\bar \theta$ in the $c_1$ and $c_5$ terms in Eq.~\eqref{eq:LO_piN}. A quick calculation shows that the $c_1$ term is again proportional to $\mathrm{Tr}(U^\dagger -U)=0$. The $c_5$ term does lead to a non-vanishing interaction
\begin{equation}\label{g0theta}
\mathcal L_{\pi N}(\bar \theta) = \bar g_0(\bar \theta)\,\bar N \vec \tau\cdot \vec \pi N\,,\qquad \bar g_0 (\bar \theta) = -\frac{1}{F_\pi} \frac{m_\star}{m_d-m_u} \delta m_N\,.
\end{equation}
This derivation reproduces the famous result of \cite{Crewther:1979pi} and shows how the $\bar \theta$ term induces an isoscalar CP-odd pion-nucleon interaction. In addition, the a priori unknown LEC $\bar g_0(\bar \theta)$ depends on the strong part of the proton-neutron mass splitting which can be determined by lattice QCD (\cite{FlavourLatticeAveragingGroupFLAG:2024oxs}) or dispersive methods (\cite{Cottingham:1963zz}). While this is a tree-level derivation, the connection to the nucleon mass splitting survives loop corrections (\cite{deVries:2015una}). Quantitative values of the LECs are discussed in Sect.~\ref{sec:NDA}. 

In similar fashion it is possible to construct CP-odd nucleon-photon and nucleon-nucleon interactions. The former give rise to direct contributions to the nucleon EDM that are necessary to renormalize loop contributions involving $
\bar g_0$ (see Sect.~\ref{sec:nucleonEDM})
\begin{equation}
  \mathcal{L}_{N\gamma}(\bar \theta)
  =
  -\,2\,\bar N\,(\bar d_0(\bar \theta) + \bar d_1(\bar \theta)\tau_3)\,S^\mu N\,v^\nu F_{\mu\nu}\,,
  \label{eq:LNgamma}
\end{equation}
where both isospin-breaking and -conserving terms appear because the quark charges break isospin. 
The nucleon-nucleon terms for the  $\bar \theta$ again conserve isospin and, ignoring terms with additional pions, take the form
\begin{equation}
\mathcal L_{NN}(\bar \theta) = \frac{1}{F_\pi \Lambda_\chi^2}\left[\bar C_1(\bar \theta) \bar N N\,\partial_\mu (\bar N S^\mu N) + \bar C_2(\bar \theta) \bar N \tau^a N\,\partial_\mu (\bar N S^\mu\,\tau^a  N)\right]\,,
\end{equation}
but are expected by naive dimensional analysis (see Sect.~\ref{sec:NDA}) to be suppressed compared to pion-exchange diagrams involving $\bar g_0(\bar \theta)$.
The nucleon-nucleon interactions are discussed in more detail in Sect.~\ref{renorm}. 

\subsection{Other CP-odd sources}
\label{sec:other_spurions}
The advantage of the spurion method is that it generalizes to the other CP-odd sources. I will go briefly through each class in turn, highlighting the key differences from the $\bar\theta$ case.\\
\\
\noindent
\textbf{Quark chromo-electric dipole moments.} The qCEDMs transform in the same chiral representation as the quark mass matrix, $(2_L, \bar 2_R) \oplus (\bar 2_L, 2_R)$, so the spurion analysis proceeds identically to the $\bar\theta$ case. One important difference is that the qCEDMs are not connected through a chiral rotation to quark masses and thus the LECs are not connected to hadron mass spectrum. A second difference is that the qCEDM have an isospin-breaking component for $\bar d_u \neq \bar d_d$. Despite these differences, the construction of the chiral Lagrangian follows that of the $\bar \theta$ term. We introduce a new spurion 
\begin{equation}
\tilde \chi = 2i\, \tilde B\,\begin{pmatrix} \tilde d_u & 0 \\ 0 & \tilde d_d \end{pmatrix}
\end{equation}
and replace $\chi \rightarrow \tilde \chi$ and $c_{1,5}\rightarrow \tilde c_{1,5}$ in the chiral Lagrangian. In the pionic sector, the isospin-breaking component, $\sim (\tilde d_u - \tilde d_d)$ leads to the appearance of pion tadpoles $\pi^0$ and related interactions between an odd number of pions. A tadpole signals an instability of the vacuum and can be eliminated through a procedure called vacuum alignment (\cite{Baluni:1978rf}). One performs an axial quark field redefinition at the quark level to ensure the vacuum is aligned with the true ground state, which eliminates tadpoles at the hadronic level. Alternatively, one can perform a pion field redefinition at the hadronic level (\cite{Mereghetti:2010tp}), but this must be accompanied by the corresponding nucleon field transformation, or spurious contributions to the nucleon couplings are generated. Either way, for the qCEDM vacuum alignment eliminates pion tadpoles and the associated multi-pion interactions.  

Because of isospin breaking, the $\bar c_1$ term now leads to a non-vanishing structure 
\begin{equation}
\mathcal L_{\pi N}(\tilde d_q) = \bar g_0(\tilde d_q)\,\bar N \vec \tau\cdot \vec \pi N + \bar g_1(\tilde d_q)\,\bar N \pi^0 N\,,\qquad \bar g_0 (\tilde d_q) = -\frac{2 \tilde c_5 \tilde B}{F_\pi} (\tilde d_u + \tilde d_d),,\qquad \bar g_1 (\tilde d_q) = -\frac{4 \tilde c_1\tilde B}{F_\pi} (\tilde d_u - \tilde d_d)\,.
\end{equation}
Nonperturbative calculations are necessary to determine the new LECs $\tilde B$ and $\tilde c_{1,5}$, but this simple derivations shows that 
 qCEDM generates both $\bar g_0$ and $\bar g_1$ at leading order, and $\bar g_1$ vanishes in the isospin limit $\bar d_u = \bar d_d$. The CP-odd nucleon-photon operators are similar to those of the $\bar \theta$ term. The CP-odd nucleon-nucleon interactions are again expected at higher order compared to pion exchange and now include two isospin-breaking terms
 \begin{equation}
\mathcal L_{NN}(\tilde d_q) = \frac{1}{F_\pi \Lambda_\chi^2}\left[\bar C_1(\tilde d_q) \bar N N\,\partial_\mu (\bar N S^\mu N) + \bar C_2(\tilde d_q) \bar N \tau^a N\,\partial_\mu (\bar N S^\mu\,\tau^a  N)+\bar C_3(\tilde d_q) \bar N N\,\partial_\mu (\bar N S^\mu \tau^3 N) + \bar C_4(\tilde d_q) \bar N \tau^3 N\,\partial_\mu (\bar N S^\mu  N)\right]\,.
\end{equation}
\\
\noindent
\textbf{Quark electric dipole moments.} The qEDMs share the same chiral representation as the qCEDMs and hence the same spurion structure. The key physical difference is that the qEDM operator contains an explicit photon field $F_{\mu\nu}$. This means that at leading order in the chiral expansion, the qEDM generates the nucleon EDM directly through the short-distance LECs $\bar d_{0,1}$ in Eq.~\eqref{eq:LNgamma}, while the interactions without photons are suppressed by $\alpha_{\rm em}$. The qEDM is therefore predominantly a ``direct'' source of nucleon EDMs rather than a source of CP-odd nuclear forces.\\
\\
\noindent
\textbf{Weinberg operator and chiral-invariant four-quark operators.} These operators are chiral singlets and the corresponding spurion has a trivial chiral transformation. The key consequence is that no CP-odd pion-nucleon operator can be built at leading order in the chiral expansion. One always needs at least one insertion of the quark-mass spurion $\chi$ or two extra derivatives and the resulting CP-odd pion-nucleon interactions are therefore suppressed by $m_\pi^2/\Lambda_\chi^2$ relative to those from the $\bar\theta$ term. This suppression implies that the nucleon EDMs operators and isospin-conserving short-range nucleon-nucleon interactions $(\bar C_{1,2}(C_W)$), which do not require pion-nucleon couplings, play a comparatively larger role than for the CP-odd chiral-breaking sources. \\
\\
\noindent
\textbf{The four-quark left-right operator.} The FQLR transforms as $(3_L, 3_R)$ under $\mathrm{SU}(2)_L \times \mathrm{SU}(2)_R$, which is a higher-dimensional representation than the quark mass and a separate construction is necessary. The detailed procedure is spelled out in (\cite{deVries:2012ab}, \cite{Bsaisou:2014oka}) and shows the appearance of a leading-order tadpole coefficient accompanied by a CP-odd three-pion vertex\begin{equation}\label{L3pi}
\mathcal L_\pi = m_N\bar \Delta\, \pi_3 \pi^2 + \dots\,.
\end{equation}
Vacuum alignment again eliminates the tadpole but in this case leaves behind the three-pion interaction. In the pion-nucleon sector the $\bar g_1$ coupling is expected and significantly larger than $
\bar g_0$, exactly opposite as for the $\bar \theta$ term. The nucleon-photon and nuclear-nucleon interactions are similar to that of the qCEDM.\\
\\
\noindent \textbf{Electron-quark operators.} The hadronization of the electron-quark operators in Eq.~\eqref{Lsl} is straightforward as the quark bilinears transform in similar fashion as the quark masses under chiral symmetry. Since the electron-structure is not affected, the leading chiral interactions are given by (\cite{Dekens:2018bci})
\begin{equation}\label{slh}
\mathcal L_{eN} = \frac{G_F}{\sqrt{2}} \left\{ \bar e i \gamma^5 e\, \bar N\left(C_S^{(0)}+C_S^{(1)}\tau^3 \right)N - 4 \bar e \sigma_{\mu\nu} e\,\bar N \left(C_T^{(0)}+C_T^{(1)}\tau^3 \right)v^\mu S^\nu N  + \bar e e\,\frac{\partial_\mu
}{m_N}\left[ N \left(C_P^{(0)}+C_P^{(1)}\tau^3 \right) S^\mu N \right]\right\}\,.
\end{equation}
The $C_S$ structures involve the electron spin an contribute to paramagnetic atomic and molecular EDMs. In fact, in the SM itself, paramagnetic EDMs are dominated by the $C_S$ contribution (\cite{Ema:2022yra}), while the electron EDM contribution is smaller by several orders of magnitude (\cite{Hoogeveen:1990cb,Yamaguchi:2020eub}). 
The $C_{P}$ interaction is suppressed in the non-relativistic limit and will be neglected below. $C_T$ require a nonzero nuclear spin and mainly contribute to EDMs of diamagnetic atoms such as ${}^{199}$Hg. 

\subsection{Values of CP-odd low-energy constants}\label{sec:NDA}

In the above sections, we saw how chiral symmetry is very useful in deriving the form of the low-energy CP-violating hadronic interactions. However, this procedure did not provide information about the values of the accompanying LECs. The success of $\chi$EFT depends on LECs to follow an expected scaling, otherwise it is difficult to see how higher-order terms can be treated in perturbation theory. This expected scaling can be derived using a technique called Naive Dimensional Analysis (NDA) introduced in \cite{Manohar:1983md}. The NDA rules are most easily summarized by using ``reduced" coupling constants (\cite{Weinberg:1989dx}) to match operators at the quark-gluon level to the hadronic level. A coupling constant $c$ of an interaction of dimension $D$ involving $N$ fields has a reduced coupling 
\begin{eqnarray}
c^R = \Lambda_\chi^{D-4}(4\pi)^{2-N} c,
\end{eqnarray}
where $\Lambda_\chi \sim 1$ GeV is the matching scale. The NDA rule is that the reduced coupling of an operator below $\Lambda_\chi$ is given by the product of the reduced couplings of the operators above $\Lambda_\chi$ that induce the operator. For example, take the CP-odd pion-nucleon interaction $\bar g_0$ with reduced coupling
$(\bar g_{0})^R = \bar g_{0}/(4\pi)$. If $\bar \theta$ is the underlying CP-violating mechanism than the NDA rules dictate that this equals  $(m_\star \bar \theta)^R = m_\star \bar \theta/\Lambda_\chi$. Rearranging this gives 
\begin{equation}
|\bar g_0 (\bar \theta)| = \mathcal O\left( \bar \theta\,m_\star \frac{(4\pi)}{\Lambda_\chi} \right) = O\left( \bar \theta\,m_\star \frac{1}{F_\pi} \right)\simeq 10^{-2}\,\bar \theta
\end{equation}
 To induce $\bar g_1(\bar \theta)$ an extra source of isospin-breaking is needed which brings in the reduced coupling $(\delta m_q)^R = (\delta m_q)/\Lambda_\chi $ where $\delta m_q = m_d-m_u$, and  NDA thus predicts $|\bar g_1(\bar \theta)| \ll |\bar g_0 (\bar 
 \theta)|$. For $\bar \theta$ it is possible to do better than NDA by using Eq.~\eqref{g0theta} and the lattice-QCD value of $\delta m_N$ (\cite{FlavourLatticeAveragingGroupFLAG:2024oxs}) to obtain 
\begin{equation}
\bar g_0 (\bar \theta)= -(17.2\pm 2.0)\cdot 10^{-3}\,\bar \theta
\end{equation}
in good agreement with the NDA estimate. A resonance saturation estimate (\cite{Bsaisou:2012rg}) gives $\bar g_1(\bar \theta) \simeq 3 \cdot 10^{-3}\,\bar \theta$ somewhat larger than NDA predicts. The same procedure can be applied to other CP-violating sources. 
For example, NDA predicts for the qCEDMs
\begin{equation}
\bar g_0 (\tilde d_q) = \mathcal O\left( (\tilde d_u + \tilde d_d)\Lambda_\chi\right)\,,\qquad \bar g_1 (\tilde d_q) = \mathcal O\left( (\tilde d_u - \tilde d_d)\Lambda_\chi\right)
\end{equation}
in decent agreement with more advanced calculations from QCD sum rules (\cite{Pospelov:2001ys})
\begin{equation}
\bar g_0(\tilde d_q) = (1 \pm 2)(\tilde d_u + \tilde d_d)\,\mathrm{GeV}\,,\qquad  \bar g_1(\tilde d_q) = (7 \pm 5)(\tilde d_u - \tilde d_d)\,\mathrm{GeV}\,,
\end{equation}
considering the large uncertainties. It is crucial to improve the determination of $\bar g_{0,1}$ arising from the qCEDMs and other sources in order to optimally interpret the outcome of EDM experiments. 

The above examples show that NDA is a reasonable, but crude, method to determine the (relative) sizes of CP-violating LECs. The NDA estimates for all relevant LECs are shown in Table~\ref{tab:NDA} and can be used as a guide to determine which interactions to include for which source. Entries labelled in dark green appear at leading order in the calculation of nucleon or nuclear CP violation, while the black entries appear at higher order. The orange entries are suppressed according to NDA, but NDA estimates are not always reliable for nucleon-nucleon interactions. This subtlety is discussed in Sect.~\ref{renorm}.  Ideally the NDA estimates are replaced by more accurate calculations in the future. The status of nonperturbative calculations of the nucleon EDMs is briefly discussed in the next section, but also in this case the NDA estimates for $\bar d_{0,1}$ turn out to be reasonable. 

For the semi-leptonic interactions the situation is better and NDA is not necessary. The LECs are given by
\begin{equation}
C_S^{(0)} = \frac{\sqrt 2}{G_F}\frac{2\sigma_{\pi N}}{m_u + m_d} c_S^{(0)}\,,\qquad C_S^{(1)} = \frac{\sqrt 2}{G_F} \frac{2\delta m_N}{m_d - m_u}  c_S^{(1)}\,,
\end{equation}
where $\sigma_{\pi N} = (59.6\pm 7.4)$ MeV (\cite{Gupta:2021ahb}) is the pion-nucleon sigma term (related to $c_1$) and $\delta m_N= (2.5 \pm 0.2)$ MeV (\cite{FlavourLatticeAveragingGroupFLAG:2024oxs}) the strong proton-neutron mass splitting (related to $c_5$). The tensor couplings are given by 
\begin{equation}
C_T^{(0)} = \frac{\sqrt 2}{G_F}(g_T^u + g_T^d)c_T^{(0)}\,,\qquad C_T^{(1)} = \frac{\sqrt 2}{G_F}(g_T^u - g_T^d)c_T^{(0)}\,,
\end{equation}
where $g_T^u \simeq -(0.21\pm0.01)$ and $g_T^d \simeq 0.82 \pm 0.03$ are the nucleon tensor charges obtained from lattice QCD (\cite{Gupta:2018lvp, FlavourLatticeAveragingGroupFLAG:2024oxs}). 

\subsection{Intermediate summary} The main message of this section is that EDMs of nucleons, nuclei, atoms, and molecules can be expressed in a relatively small number of LECs
\begin{equation}\label{eq:LECs}
  \{\bar \Delta, \bar g_0,\,\bar g_1,\bar d_0,\,\bar d_1,\,\bar C_{i},\,C_S^{(0,1)}
  ,\,C_T^{(0,1)}\}\,.
\end{equation}
Depending on the CP-violating source under consideration a different subset of these interactions is expected to play a role. 

The uncertainties on these LECs vary widely across the different sources 
and represent one of the main bottlenecks in the interpretation of EDM 
experiments. For the $\bar\theta$ term, the situation is relatively 
favorable as $\bar g_0(\bar\theta)$ is tied to the strong 
neutron-proton mass splitting determined accurately by lattice QCD. 
The semi-leptonic LECs $C_S^{(0,1)}$ and $C_T^{(0,1)}$ originating from semi-leptonic four-fermion operators are similarly 
well-determined, being related to $\sigma_{\pi N}$, $\delta m_N$, and 
the nucleon tensor charges, all of which are now precisely known from 
lattice QCD. For the qCEDMs,  the situation is much less 
satisfactory. The QCD sum rule estimates for $\bar g_{0,1}(\tilde d_q)$ 
carry $\mathcal{O}(100\%)$ uncertainties, and no lattice QCD calculations 
currently exist. The LECs for the Weinberg operator and chiral-invariant 
four-quark operators are even less constrained, with only order-of-magnitude estimates available. These uncertainties propagate directly into the interpretation of EDMs limiting the usefulness of EDM 
measurements in constraining and hopefully identifying these sources. Improving the determination of these LECs through lattice QCD, see \cite{Liu:2024kqy} for a recent review, 
should therefore be a high priority.  

A promising path forward is provided by the 
gradient flow (\cite{Luscher:2010iy}) which 
offers a gauge-invariant and systematically improvable renormalization 
scheme for higher-dimensional CP-odd operators on the lattice. The 
perturbative matching of the qCEDMs, Weinberg operator, and the CP-odd four-quark 
operators to the gradient-flow scheme has recently been completed at 
one loop~(\cite{Cirigliano:2020msr, Mereghetti:2021nkt,Buhler:2023gsg}) laying the 
groundwork for lattice QCD determinations of the 
hadronic LECs. Preliminary calculations have appeared (\cite{Kim:2021qae, Bhattacharya:2025aht}), see \cite{Shindler:2021bcx} for a more comprehensive discussion.

\begin{table}[t]
\centering
\renewcommand{\arraystretch}{1.6}
\newsavebox{\mytable}
\begin{lrbox}{\mytable}
\setlength{\tabcolsep}{15pt}
\begin{tabular}{||c||ccccc||}
\hline
Source 
  & $\bar{\theta}$ 
  & qCEDM 
  & FQLR 
  & qEDM 
  & Weinberg \\
\hline\hline
$\bar{g}_0$
  & $\color{darkgreen}{\bar{\theta}\,(m_\star/F_\pi)}$
  & $\color{darkgreen}{\tilde{d}_q\,\Lambda_\chi}$
  & $C_{\mathrm{LR}}\,m_q\,F_\pi$
  & $d_q\,\Lambda_\chi\,\dfrac{\alpha_{\mathrm{em}}}{4\pi}$
  & $C_W\,m_q\,\Lambda_\chi$ \\
$\bar{g}_1/\bar{g}_0$
  & $m_q/\Lambda_\chi$
  & $\color{darkgreen}{1}$
  & $\color{darkgreen}{\Lambda_\chi/m_q}$
  & $1$
  & $1$ \\
\hline
$\bar{\Delta}/\bar{g}_0$
  & $m_q/\Lambda_\chi$
  & $m_q/\Lambda_\chi$
  & $\color{darkgreen}{\Lambda_\chi/m_q}$
  & $1$
  & $m_q/F_\pi$ \\
\hline
$\bar{d}_{0,1}$
  & $\color{darkgreen}{e\,\bar{\theta}\,(m_\star/\Lambda_\chi^2)}$
  & $\color{darkgreen}{e\,\tilde{d}_q\,(F_\pi/\Lambda_\chi)}$
  & $\color{darkgreen}{e\,C_{\mathrm{LR}}\,(F_\pi^2/\Lambda_\chi)}$
  & $\color{darkgreen}{d_q}$
  & $\color{darkgreen}{e\,C_W\,F_\pi}$ \\
\hline
$\bar{C}_{1,2}$
  & $\color{orange}{\bar{\theta}\,(m_\star/F_\pi)}$
  & $\color{orange}{\tilde{d}_q\,\Lambda_\chi}$
  & $C_{\mathrm{LR}}\,m_q\,F_\pi$
  & $d_q\,\Lambda_\chi\,\dfrac{\alpha_{\mathrm{em}}}{4\pi}$
  & $\color{darkgreen}{C_W\,\Lambda_\chi^2}$ \\
$\bar{C}_{3,4}/\bar{C}_{1,2}$
  & $m_q/\Lambda_\chi$
  & $\color{orange}{1}$
  & $\color{orange}{\Lambda_\chi/m_q}$
  & $1$
  & $m_q/\Lambda_\chi$ \\
\hline
\end{tabular}
\end{lrbox}
\usebox{\mytable}
\captionsetup{width=\wd\mytable}
\caption{NDA estimates of the CP-odd hadronic LECs for each hadronic source of CP violation.
Entries in \textcolor{darkgreen}{dark green} indicate leading-order contributions,
while entries in \textcolor{orange}{orange} denote contributions that are subleading by NDA but could  possibly be enhanced through renormalization, see Sect.~\ref{renorm}. Uncolored entries are suppressed by at least one power
of $m_q/\Lambda_\chi$ relative to the leading contribution in the same
column. Here $m_\star = m_u m_d/(m_u+m_d)$ is the reduced quark mass, $m_q$ denotes the quark mass, 
$\Lambda_\chi \sim 1$~GeV is the chiral symmetry breaking scale, and
$F_\pi \simeq 92.2$~MeV is the pion decay constant.}
\label{tab:NDA}
\end{table}

\section{Electric dipole moments of nucleons and nuclei}

\subsection{Nucleon electric dipole moments and Schiff moments}\label{sec:nucleonEDM}
The EDMs of the neutron and proton provide the most direct measurements of hadronic CP violation without requiring further atomic/molecular calculations of screening or enhancement factors.  That being said, the computation of nucleon EDMs in terms of the underlying CP-odd mechanism is not an easy task. Various approaches exist in the literature including quark models (\cite{Dib:2006hk,Yamanaka:2020kjo}), QCD sum rules (\cite{Pospelov:1999ha,Hisano:2012sc,Haisch:2019bml}), $\chi$PT (\cite{Borasoy:2000pq,Hockings:2005cn,deVries:2010ah}), holography (\cite{Hong:2007tf,Bartolini:2016jxq}), and lattice QCD.  Most effort has focused on the QCD $\bar \theta$ term and the quark EDMs, while much less is know about the quark chromo-EDMs, Weinberg operator, or four-quark interactions. 

Chiral techniques are useful when the nucleon EDMs depend on a chiral logarithm $\sim \log m_\pi$ which is enhanced in the chiral limit. However, these contributions are typically divergent and thus require short-distance (here short-distance means from distances shorter-than-pion-range) contributions that renormalize the EDM. $\chi$PT does not predict the values of these short-distance contributions leading to an increases theoretical uncertainty. One-loop diagrams involving $\bar g_0$ and $\bar g_1$ have been computed up to next-to-leading order (\cite{Crewther:1979pi,Ottnad:2009jw,Mereghetti:2010kp}) and give 
\bea \label{eq:dnpchiral}
  d_n& = & {\bar d}_0(\mu) -\bar d_1(\mu)-\frac{e g_A\bar g_0}{8\pi^2 F_\pi} \left(  \ln
\frac{m_\pi^2}{\mu^2} -\frac{\pi m_\pi}{2 m_N} \right)\,\,\,,\nonumber\\
d_p & = & {\bar d}_0(\mu) + \bar d_1(\mu)+\frac{e g_A}{8\pi^2 F_\pi} \left[ \bar g_0 \left(  \ln
\frac{m_\pi^2}{\mu^2} -\frac{2\pi m_\pi}{m_N} \right) -
\bar g_1 \frac{\pi m_\pi}{2 m_N} \right]\,\,\,.
\eea
The short-distance contributions $\bar d_0\mp \bar d_1$ have a scale dependence in order to absorb the scale dependence of the chiral logarithms. At this order, the neutron EDM does not depend on $\bar g_1$ and the first dependence appears at next-to-next-to-leading order (\cite{Seng:2014pba}). Setting $\mu=m_N$ and assuming the chiral logarithm to dominate the short-distance pieces gives
\begin{equation}
d_n(\bar \theta) \simeq - d_p(\bar \theta) \simeq 10^{-3}\,\bar \theta\,e\,\mathrm{fm}\,,
\end{equation}
in good agreement with results from QCD sum rules (\cite{Pospelov:1999ha}). More recently lattice QCD (\cite{Dragos:2019oxn,Liang:2023jfj}) has been applied to compute the nucleon EDMs, see \cite{Liu:2024kqy} for a   review, giving 
\bea \label{nucleonEDM}
  d_n(\bar \theta)  =  -(1.48 \pm0.34)\cdot 10^{-3}\,\bar \theta\,e\,\mathrm{fm}\,,\qquad d_p(\bar \theta)  =  (3.8 \pm 1.4)\cdot 10^{-3}\,\bar \theta\,e\,\mathrm{fm}\,.
\eea
For the quark chromo-EDM the pion loops are also expected to be dominant but the values of $\bar g_{0,1}$ are less well known. QCD sum rules results are obtained in (\cite{Pospelov:2000bw,Hisano:2012sc}).

For quark EDMs, the pion-loops are subleading and chiral techniques are not useful. In this case, lattice QCD calculations of nucleon tensor charges have become very accurate (\cite{Gupta:2018lvp}, \cite{FlavourLatticeAveragingGroupFLAG:2024oxs}) and 
\bea
d_n(d_q) = g_T^u d_u + g_T^d d_d\,,\qquad  d_p(d_q) = g_T^d d_u + g_T^u d_d\,,
\eea
where the tensor charges are given in Sect.~\ref{sec:NDA}. The role of the strange quark EDM is less clear.  For chiral-symmetry-conserving sources such as the Weinberg operator, the pion loops are expected to be subleading as well. In this case only estimates exist based on QCD sum rules (\cite{Demir:2002gg,Haisch:2019bml}) and quark models (\cite{Yamanaka:2020kjo}) giving
\begin{equation}
d_n(C_W) \simeq (30\pm 20) \,\mathrm{MeV}\,e\,C_W\,,
\end{equation}
where the uncertainty is chosen to span the range of predictions. The proton EDM comes with a similar matrix element but with opposite sign (\cite{Haisch:2019bml}). These values are consistent with the NDA expectation of Table~\ref{tab:NDA}.

While chiral techniques are of limited use for nucleon EDMs, the momentum dependence encoded in the nucleon electric dipole form factor (EDFF) can be predicted. The nucleon EDFF can be decomposed 
as
\begin{equation}\label{EDFF}
F_{n,p}(Q^2) = d_{n,p} - S'_{n,p} Q^2 + H_{n,p}(Q^2)\,,
\end{equation}
where $Q^2 = -q^2 >0$ indicates the momentum transfer from the photon with outgoing four-momentum $q^\mu$. For chiral-breaking CP-odd sources, the nucleon Schiff moments, $S'_{n,p}$ (the prime indicates that this is not the full nucleon Schiff moment as discussed in Sect.~\ref{sec:dia}), are dominated by the pion cloud (\cite{Thomas:1994wi}) and, up to small isospin-breaking corrections, isovector in nature (\cite{Mereghetti:2010kp})
\begin{equation}\label{nucleonschiff}
S'_n = - S'_p = \frac{e g_A \bar g_0}{48 \pi^2 F_\pi m_\pi^2}\left(1-\frac{5 \pi m_\pi}{4 m_N}\right)\,.
\end{equation}
The EDFF shape functions encoded in $H_{n,p}$ start at $\mathcal O(Q^4)$ and are also completely specified (\cite{deVries:2010ah}). They can be used to guide lattice QCD extrapolations in the $Q^2\rightarrow 0$ limit. For the qEDM and chiral-invariant sources, the nucleon Schiff moments are suppressed and come with undetermined low-energy constants. This implies that the ratio of nucleon Schiff moments to nucleon EDMs is indicative of the underlying source of CP violation. Unfortunately nucleon Schiff moments are not directly measured and probing this ratio is difficult. 

\subsection{Electric dipole moments of atomic nuclei}

After the nucleon EDMs, the next level in complexity is to consider nuclear EDMs. Up to very recently, no nuclear EDMs were directly measured but this changed with the first limit on the deuteron EDM set by the JEDI collaboration (\cite{Andres:2026lho}). By storing the deuteron $(D)$ in an electromagnetic storage ring and tracking the tilt of the spin axis with respect to the ring plane, it was possible to constrain
\begin{equation}
d_D < 2.5\cdot 10^{-17}\,e\,\mathrm{cm}\,.
\end{equation}
While this limit is 9 orders away from the direct neutron EDM limit, it provides a proof-of-principle measurement and motivates the construction of future dedicated experiments. Such storage rings have been proposed for protons, light nuclei, light ions, and muons (\cite{pEDM:2022ytu}, \cite{Adelmann:2025nev}, \cite{Dutsov:2025kbd}).

From the theoretical point of view, the main difference between nuclear and nucleon EDMs is the contribution from multi-nucleon CP-odd mechanisms.  In addition to the single nucleon EDM contributions, there are novel contributions from multi-nucleon CP-odd electromagnetic currents and from the interplay of CP-odd nuclear forces and CP-conserving currents. Because nuclear calculations are difficult (even for the deuteron) it is useful to use power counting to assess the (relative) sizes of the various contributions. 

\begin{figure}[t!]
    \centering
    \includegraphics[width=0.8\textwidth]{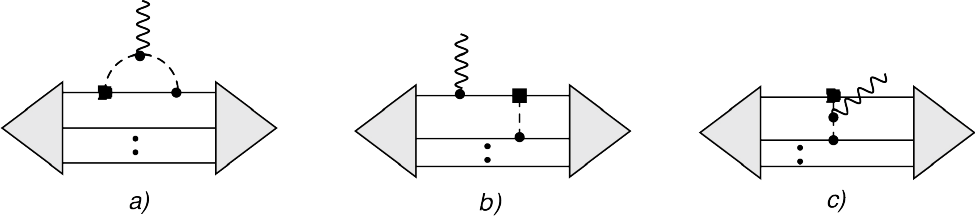}
    \caption{Three contributions to a nuclear EDM described in the text. Solid, wavy, and dashed lines denote nucleons, photons, and pions, respectively. The triangle denotes the nuclear wavefunction and the dots denote the $A-2$ nucleon propagators. The black square denotes the CP-odd pion-nucleon vertex $\bar g_0$ while the black circles denote CP-conserving interactions.}
    \label{fig:classes}
\end{figure}

Let us consider three contributions shown in Fig.~\ref{fig:classes} involving the CP-odd pion-nucleon vertex $\bar g_0$. Diagram \ref{fig:classes}a) represents a one-loop contribution to the nucleon EDM (see Eq.~\eqref{eq:dnpchiral}) which then enters the nuclear system, diagram \ref{fig:classes}b) represents an insertion of a CP-odd pion exchange involving $\bar g_0$ and $g_A$ and an insertion of the proton charge, while finally diagram \ref{fig:classes}c) is a contribution from a CP-odd pion-in-flight current involving one $\bar g_0$ vertex and one strong CP-conserving pion-nucleon vertex $\sim g_A$.

We now briefly discuss the power counting of these diagrams. In heavy-baryon  $\chi$PT, in loop diagrams involving a single nucleon field, it is always possible to avoid the nucleon pole when doing the virtual $k^0$ energy integration. In such diagrams, the pion and nucleon propagators count respectively as $1/Q^2$ and $1/Q$ where $Q\sim m_\pi$. In addition, each loop integration counts as $Q^4/(4\pi)^2$. Weinberg noticed that in two-nucleon diagrams the nucleon poles cannot always be avoided leading to a `pinch' singularity (\cite{Weinberg:1990rz}) and picking up energies $\sim Q^2/m_N$ instead. This enhances each nucleon propagator to count as $m_N/Q^2$ instead of $1/Q$. The integration measure now counts as $Q^5/(4\pi m_N)$ where the extra $4\pi$ arises from the loop topologies (\cite{vanKolck:2020plz}). Using these rules we can now quickly estimate the contributions from the 3 diagrams in Fig.~\ref{fig:classes}. We normalize the diagrams by omitting the common $A-1$ loop integrations and $A+1$ nucleon propagators that contribute to all diagrams. Diagram a) then simply contributes 
\begin{equation}
D_a = \mathcal O( Q\, d_{n,p}) = \mathcal O\left(\frac{e g_A \bar g_0}{(4\pi)^2 } \right) \,,
\end{equation}
where the $Q$ arises from the derivative in the nucleon EDM vertex. The second equality uses $Q \sim F_\pi$ and holds for sources where the nucleon EDMs get leading contributions from pion loops. Now consider diagram b) which contains one extra loop, $ Q^5/(4\pi m_N)$, two extra nucleon propagators, $m_N^2/Q^4$, one pion propagator, $1/Q^2$, and a combination of pion-nucleon vertices $Q g_A \bar g_0/F_\pi \sim g_A \bar g_0$. Combining the factors predicts
\begin{equation}\label{PCDb}
D_b = \mathcal O \left( e \frac{g_A \bar g_0 m_N}{4\pi Q}\right)\,.
\end{equation}
Using $4\pi Q \sim m_N$ shows that diagram b), arising from CP-odd nuclear forces, is enhanced by $(4\pi)^2 \sim \Lambda_\chi^2/Q^2$ with respect to the nucleon EDM contributions. Such an enhancement was already noted a long time ago in \cite{Flambaum:1984fb}.

Contribution from CP-odd two-nucleon currents are bit trickier to count. The topology of diagram $c)$ involves a pion-photon vertex which, in order to generate an EDM, contributes a pion energy $q^0 \sim Q^2/m_N$. Taking this suppression into account leads to 
\begin{equation}
D_c = \mathcal O\left(\frac{e g_A \bar g_0}{(4\pi)^2 } \right) \,,
\end{equation}
at the same order as the nucleon EDMs. Note that at this order several other CP-odd currents appear (\cite{deVries:2011an}). 

Of course, power counting only provides a guide and explicit calculations are necessary to confirm the above results, but the initial conclusion is that EDMs (and Schiff moments) of nuclear systems are dominated by CP-odd nuclear forces that cause nuclear ground states to obtain a small admixture of opposite parity states. The CP-conserving electromagnetic current connects the parity-admixed states back into the ground state. Schematically, the calculation involves calculating the nuclear ground state wave function, $\Psi_A$, by solving a Schr\"odinger equation involving a CP-even nuclear Hamiltonian, $\mathcal H_{CP}$, and then perturbing this wave function with a CP-odd nuclear potential, $V_{\CPv}$, to obtain the parity-admixed wave function $\tilde \Psi_A$:
\begin{eqnarray}
(E-\mathcal H_{CP})|\Psi_A\rangle &=&0\,,\nonumber\\
(E-\mathcal H_{CP})|\tilde \Psi_A\rangle &=& V_{\CPv}\, |\Psi_A\rangle\,.
\end{eqnarray}
The EDM is then proportional to the transition matrix element 
\begin{equation}
d_A \sim \langle \Psi_A | J_{CP}| \tilde \Psi_A \rangle\,,
\end{equation}
where $J_{CP}$ is the CP-conserving current which, at leading order, simply arises from minimal coupling to the proton charge but gets higher-order corrections, for example, from photons coupling to pions-in-flight. 

\subsection{The CP-violating nucleon-nucleon potential}
The above discussion shows that the calculation of nuclear EDMs and Schiff moments requires the CP-odd nucleon-nucleon (or multi-nucleon potential). Historically this potential was derived using one-meson-exchange models involving a combination of CP-even and CP-odd meson-nucleon interactions. The CP-violating potential can also be calculated using $\chi$EFT which has the main advantage that higher-order corrections, not necessarily described by one-meson exchange, can be computed systematically. A detailed review of the derivation of CP-violating nuclear potentials is given by \cite{deVries:2020iea}, and here I discuss the main features.

\begin{figure}[t!]
    \centering
    \includegraphics[width=0.8\textwidth]{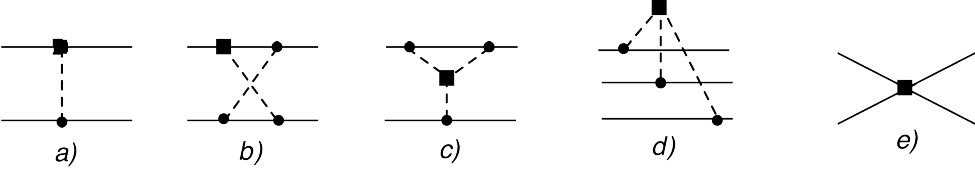}
    \caption{Several diagrams contributing to the CP-odd nucleon-nucleon potential. Solid and dashed lines denote nucleons and pion, respectively. Black squares (black circles) are CP-odd (CP-even) interactions. }
    \label{fig:potential}
\end{figure}

CP-violating nuclear forces  can be computed from the interactions in Sect.~\ref{sec:chiralEFT}. The pion-nucleon, three-pion vertices, and short-range interactions contribute to nucleon-nucleon interactions through diagrams depicted in Fig.~\ref{fig:potential}. 
The most important diagrams are the one-pion-exchange (OPE) diagrams, Fig.~\ref{fig:potential}a),  which lead to 
\begin{equation}
V^{\mathrm{OPE}}_{\CPv}= i\frac{g_A\bar g_0}{2 F_\pi}(\vec \tau_1\cdot \vec \tau_2)\frac{\vec k \cdot (\vec \sigma_1-\vec \sigma_2)}{\vec k^2 + m_\pi^2} + i\frac{g_A\bar g_1}{4 F_\pi}\left[ (\tau_1^3 + \tau_2^3)\frac{\vec k \cdot (\vec \sigma_1-\vec \sigma_2)}{\vec k^2 + m_\pi^2} + (\tau_1^3 - \tau_2^3)\frac{\vec k \cdot (\vec \sigma_1+\vec \sigma_2)}{\vec k^2 + m_\pi^2}\right] \,,
\end{equation}
where $\vec \tau_{1,2}$ and $\vec\sigma_{1,2}$ are, respectively, the isospin and spin of the involved nucleons, and $\vec k = \vec p_1'-\vec p_1$ is the momentum transfer between the nucleons.  The tree-level contributions are obtained from using standard Feynman rules and neglecting the virtual pion energies $k_0\ll |\vec k|$ as indicated by the power counting rules in the previous section. The $\bar g_0$ OPE potential conserves isospin and leads to ${}^1S_0$-${}^3P_0$ mixing with equal strengths for $nn$, $pp$, and $np$ (here $n$ and $p$ denote neutron and proton), and to ${}^3 S_1$-${}^1 P_1$ mixing for $np$. The first term in the $\bar g_1$ OPE potential however violates isospin and leads to ${}^1S_0$-${}^3P_0$ mixing for $pp$ and $nn$ (with opposite sign) but not for $np$. The second term proportional to $\bar g_1$ changes total isospin by one unit and leads to ${}^3S_1$-${}^3P_1$ for $np$ systems. 
CP-odd two-pion exchange potentials, Fig.~\ref{fig:potential}b), appear at next-to-next-to-leading order (and beyond) (\cite{Maekawa:2011vs}) and are not negligible at least for EDMs of light nuclei (\cite{Gnech:2019dod}). The role of $
\Delta$-baryons has been investigated showing that no new CP-violating LECs are required (\cite{Gandor:2024kty}).

 The three-pion vertex in Eq.~\eqref{L3pi} contributes at one loop, Fig.~\ref{fig:potential}c), to the same spin-isospin structure as $\bar g_1$ but with more complicated dependence on the momentum transfer $\vec k^2$ (\cite{deVries:2012ab})
 \begin{equation}\label{Vdelta}
V^{\mathrm{OPE},\bar \Delta}_{\CPv}=  -i\frac{15 g_A^3 m_\pi m_N \bar 
\Delta}{128 \pi F_\pi^3}\left[ (\tau_1^3 + \tau_2^3)\frac{\vec k \cdot (\vec \sigma_1-\vec \sigma_2)}{\vec k^2 + m_\pi^2} + (\tau_1^3 - \tau_2^3)\frac{\vec k \cdot (\vec \sigma_1+\vec \sigma_2)}{\vec k^2 + m_\pi^2}\right]\left[1+ f\left(\frac{|\vec k|}{2m_\pi}\right)\right] \,,
\end{equation}
where 
\begin{equation}
    f(x)=\frac{1+2x^2}{3x}\,\arctan(x)-\frac{1}{3}\,.
\end{equation} 
The part independent of $f(x)$ in Eq.~\eqref{Vdelta} provides a renormalization of $\bar g_1$ (\cite{deVries:2012ab}). Higher-order one-loop diagrams have been computed by \cite{Gnech:2019dod}.
The three-pion vertex also leads to a CP-odd three-nucleon potential, Fig.~\ref{fig:potential}d), at tree level
\begin{equation}\label{VNNN}
V^{\mathrm{NNN}}_{\CPv} = -i\frac{g_A^3 m_N \bar \Delta}{4 F_\pi^2} \left( \vec \tau_1 \cdot \vec \tau_2\,\tau_3^3 + \vec \tau_1 \cdot \vec \tau_3\,\tau_2^3+\vec \tau_2 \cdot \vec \tau_3\,\tau_1^3 \right)\,\frac{(\vec k_1 \cdot \vec \sigma_1)(\vec k_2\cdot \vec \sigma_2)(\vec k_3 \cdot\vec \sigma_3)}{(\vec k_1^2 + m_\pi^2)(\vec k_2^2 + m_\pi^2)(\vec k_3^2 + m_\pi^2)}\,,
\end{equation}
where $\vec k_i$ is the three-momentum that flows to each nucleon. 

The CP-violating nucleon-nucleon interactions, Fig.~\ref{fig:potential}e), give rise to short-range potentials
\begin{equation}
V^{\mathrm{short}}_{\CPv} = \frac{1}{F_\pi \Lambda_\chi^2}\left[i \bar C_1\,\vec k \cdot (\vec \sigma_1-\vec \sigma_2) + i \bar C_2\,\vec k \cdot (\vec \sigma_1-\vec \sigma_2)\vec \tau_1 \cdot \vec \tau_2 + i\frac{\bar C_3 + \bar C_4}{2} \,\vec k \cdot (\vec \sigma_1-\vec \sigma_2) (\tau_1^3 + \tau_2^3) + i\frac{\bar C_3 - \bar C_4}{2} \,\vec k \cdot (\vec \sigma_1+\vec \sigma_2) (\tau_1^3 - \tau_2^3)\right]\,.
\end{equation}

Which of the above terms are relevant depends on the underlying CP-violating source, see Table~\ref{tab:NDA}. For the $\bar \theta$ term and qCEDM the OPE potentials proportional to, respectively, $\bar g_0$ and $\bar g_{0,1}$ are expected to dominate nuclear EDMs. For the FQLR $\bar g_1$ and the induced terms by $\bar \Delta$ appear at leading order. For the Weinberg operator, on the other hand, we expect relevant contributions from both OPE as well as the short-range interactions $\bar C_{1,2}$.

\subsection{Renormalization of CP-odd nuclear forces and short-distance nucleon-nucleon interactions }\label{renorm}
The dominant CP-odd nuclear forces are usually assumed to arise from long-range one-pion exchange (OPE) generated by CP-odd pion–nucleon couplings. For chiral-symmetry-breaking sources of CP violation such as the $\bar \theta$ term and qCEDMs, power counting combined with NDA then places purely short-distance CP-odd $N\!N$ operators at next-to-next-to-leading order (see Table~\ref{tab:NDA}), suggesting that nuclear EDMs are largely controlled by a small set of long-range couplings. 
The analysis in \cite{deVries:2020loy} revisited this assumption by enforcing renormalization of CP-odd amplitudes in channels where the CP-even tensor nucleon-nucleon ($N\!N$) force is strong and attractive. In the ${}^3P_0$ partial wave the attractive tensor OPE is singular (\cite{Nogga:2005hy}) and this leads to ${}^3P_0$ phase shifts that are very sensitive to the applied regulator. \cite{Nogga:2005hy} suggested to promote $P$-wave nucleon-nucleon interactions to leading order, going against NDA expectations. A much more detailed discussion can be found in \cite{vanKolck:2020llt}. 

After renormalizing the CP-conserving interactions we  add the CP-odd potential as a perturbation. When the long-range CP-odd OPE potential involving $\bar g_0$ or $\bar g_1$ is iterated together with the strong $N\!N$ interaction, the resulting CP-odd mixing amplitude for ${}^1S_0$–${}^3P_0$ transitions shows a strong and oscillatory dependence on the ultraviolet regulator. Since observable CP-odd mixing angles must be regulator independent, this behaviour signals that CP-violating OPE alone is not sufficient: a short-distance counterterm is required to absorb the divergence. 
The required counterterms have the structure of a local CP-odd $N\!N$ interaction that induces ${}^1S_0$–${}^3P_0$ transitions and are described by the combinations $\bar C_1 + \bar C_2$ (for the $\bar g_0$ OPE) and $\bar C_3+\bar C_4$ (for the $\bar g_1$ OPE). NDA would assign these short-range interactions to subleading order, but the renormalization analysis shows that its coefficient is enhanced and must be counted as leading order. This is signaled by the orange entries in Table \ref{tab:NDA}. A  similar enhancement of short-distance $N\!N$ physics occurs in the $\chi$EFT description of neutrinoless double beta decay (\cite{Cirigliano:2018hja}). Once the short-distance LECs are promoted to leading order, the theory can be renormalized and the ${}^1S_0$–${}^3P_0$ mixing amplitude becomes regulator independent for a wide range of cutoffs. Other CP-odd mixing, such as those for ${}^3S_1$–${}^1P_1$ and ${}^3S_1$–${}^3P_1$, are already stable and remain dominated by long-range OPE.

This has direct implications for EDM phenomenology. Many nuclear EDMs receive important contributions from ${}^1S_0$–${}^3P_0$ transitions in proton–neutron, proton–proton, and neutron–neutron pairs, so their values generally depend on the short-distance LEC $\bar C_1+\bar C_2$ and $\bar C_3+\bar C_4$ at the same order as on the long-range $\bar g_{0,1}$ terms. In other words, for these systems one cannot reliably predict EDMs (or their ratios) in terms of the underlying QCD $\bar\theta$ angle or other CP-odd sources using pion-exchange interactions alone. Consequently, nuclear EDMs and Schiff moments of light and heavy nuclei such as ${}^3$He, ${}^{199}$Hg, and ${}^{225}$Ra, which are sensitive to ${}^1S_0$–${}^3P_0$ mixing, would receive contribution from both long-range pions and LO short-range CP-odd $N\!N$ contact terms. An important exception is the deuteron EDM, which, as discussed below, is dominated by ${}^3S_1$–${}^3P_1$ mixing. 

A central challenge is  to determine the short-range LECs in order to assess whether they really play an important role.
Lattice QCD calculations of $N\!N$ scattering in a nonzero $\bar\theta$ (or other CPV source) background could in principle allow a direct matching of chiral EFT to QCD, though this is technically demanding. Second, for the $\bar\theta$ term, chiral symmetry relates the CP-odd contact interaction  to CP-even but isospin-breaking $NN\pi$ operators that contribute to charge-symmetry breaking in pion production reactions such as $NN\to d\pi$ (\cite{vanKolck:2000ip}) and $dd\to\alpha\pi^0$ (\cite{Nogga:2006cp}). Precision data on these processes, analyzed within renormalized chiral EFT, may thus provide an indirect handle on the short-distance CP-odd $N\!N$ couplings. Right now such studies have not been carried out, leaving the role of short-range CP-odd nuclear forces an open question.  

\subsection{Electric dipole moments of light nuclei}

We are now ready to discuss explicit computation of nuclear EDMs. The most interesting system is the deuteron which is relatively easy to describe and has been constrained experimentally (\cite{Andres:2026lho}). The deuteron EDM has been computed with various  theoretical methods using phenomenological meson-exchange CP-conserving and CP-violating potentials (\cite{Afnan:2010xd,Liu:2004tq}) to $\chi$EFT calculations (\cite{deVries:2011an, Bsaisou:2014zwa,Gnech:2019dod}) to holographic methods (\cite{Bartolini:2019shp}). To reasonable accuracy the deuteron EDM can be calculated analytically in $\chi$PT under the assumption that CP-conserving pion exchange is treated perturbatively (\cite{deVries:2011re}), a decent approximation in  a loosely bound nuclear system such as the deuteron. In this approximation, the leading  CP-conserving nucleon-nucleon potential is a contact potential leading to a zero-range  deuteron wave function and, at leading order, the EDM results agree with \cite{Khriplovich:1999qr}. In this approach, the deuteron EDM is given by 
\begin{equation}
d_D = d_n + d_p + \frac{e g_A \bar g_1}{12 \pi F_\pi}\frac{m_N}{m_\pi} \frac{1+\xi}{(1+2\xi)^2}\simeq d_n +d_p + 0.23\,\bar g_1\,e\,\mathrm{fm}\,.
\end{equation}
Here $\xi = \gamma/m_\pi \simeq 0.33$ where $\gamma = \sqrt{m_N E_b}t \simeq 45$ MeV in terms of the  deuteron binding energy $E_b = 2.2$ MeV. The analytical result explicitly confirms the power-counting expectation in Eq.~\eqref{PCDb}. The deuteron spin-isospin properties (the ground state is mainly ${}^3S_1$ with a small ${}^3D_1$ admixture) ensure that $\bar g_0$ and $\bar C_{1,2}$ do not contribute to the deuteron EDM at this order. Isospin-breaking corrections appear at higher order but are very small (\cite{deVries:2011an}). The deuteron EDM results have been confirmed with more advanced numerical calculations using phenomenological and $\chi$EFT potentials to describe the deuteron wave function (\cite{Bsaisou:2014zwa}, \cite{Gnech:2019dod}). This  slightly changes the numerical coefficients in front of the CP-odd LECs
\begin{equation}
d_D = 0.94(d_n + d_p) + \left[0.19\,\bar g_1-0.30\,\bar \Delta\right]\,e\,\mathrm{fm}\,,
\end{equation}
where the uncertainty on the $\bar g_1$ and $\bar \Delta$ coefficients is around $20\%$ based on regulator variations in the numerical calculations and missing higher-order corrections. The three-pion vertex $\sim \bar \Delta$ mainly contributes by effectively renormalizing $\bar g_1$ with roughly $10\%$ contributions from the $f(|\vec k/(2m_\pi))$ part in Eq.~\eqref{Vdelta}.

While not directly targeted in experimental storage rings, although plans do exist (\cite{pEDM:2022ytu,Dutsov:2025kbd}), it is interesting to slowly increase the number of nucleons. The ${}^3$He and ${}^3$H EDMs have been calculated with one-meson-exchange potentials (\cite{Stetcu:2008vt, Song:2012yh}), pionless EFT (\cite{Yang:2020ges}), and $\chi$EFT (\cite{deVries:2011an, Bsaisou:2014zwa, Gnech:2019dod}). Compared to the deuteron, the main difference is the sensitivity to $\bar g_0$ and $\bar C_{i}$ and to the CP-odd three-nucleon force induced by the three-pion interaction. Using the most recent results from \cite{Gnech:2019dod}, the ${}^3$He EDM is 
\begin{eqnarray}\label{d3He}
d_{{}^3\mathrm{He}} &=& 0.91\,d_n -0.033\, d_p + \left[\left(0.056\,\bar g_0 + 0.16\, \bar g_1\right) + \left(-0.20-0.18\right)\,\bar \Delta \right] \,e\,\mathrm{fm}\nonumber\\
&& + \left[ 0.002 (\bar C_1 + \bar C_2) -0.008  (\bar C_1 - \bar C_2) -0.01 (\bar C_3 - \bar C_4)  \right] \,e\,\mathrm{fm}\,.
\end{eqnarray}
The ${}^3$H EDM is obtained by swapping $d_n \leftrightarrow d_p$ and setting $\bar g_0 \rightarrow -\bar g_0$ and $\bar C_{1,2} \rightarrow -\bar C_{1,2} $. 
The ${}^3$He EDM is essentially the neutron EDM combined with contributions from the CP-odd potential. Compared to the deuteron there is a comparable sensitivity to $\bar g_1$ and a somewhat smaller dependence on $\bar g_0$. The contributions from $\bar \Delta$ is split into a loop-induced piece from Eq.~\eqref{Vdelta} and higher-order corrections ($-0.2$) and a contribution ($-0.18$) from the CP-odd three nucleon-force in Eq.~\eqref{VNNN}. The three-nucleon contribution was also computed by \cite{Bsaisou:2014zwa} and found to be smaller by an order of magnitude. The reason for the discrepancy with \cite{Gnech:2019dod} is not clear and motivates an independent calculation. The second line of Eq.~\eqref{d3He} contains contributions from the short-distance CP-odd nucleon-nucleon interactions. By NDA these are expected to small for sources like $\bar \theta$, qCEDMs, and the FQLR but for chiral-invariant sources $\bar C_{1,2}$ are expected to give relevant contributions. I have added a contribution from $\bar C_3 -\bar C_4$ which, although expected to be small by NDA, might have to be promoted to leading order to ensure proper renormalization as discussed in Sect.~\ref{renorm}. 

Calculations have been extended to larger (but still light) nuclei in various approaches. In a series of works, the $\alpha$ cluster model was applied to compute the EDMs of nuclei such as ${}^6$Li, ${}^9$Be, and ${}^{13}$C (\cite{Yamanaka:2015qfa,Yamanaka:2016umw,Yamanaka:2019vec}). The model is based on the observation that the $\alpha$ cluster is stable and larger systems can be described as collections of $\alpha$ particles and individual nucleons, simplifying the many-body problem (\cite{Yamada:2011bi}). For example, ${}^6$Li can be seen as a $\alpha$-$n$-$p$ system where the $n$-$p$ system is well described by a deuteron cluster. The ${}^6$Li EDM is then arising approximately from the deuteron EDM in addition to contributions from a $\alpha$-nucleon CP-odd potential. The latter potential is obtained by folding the CP-odd nucleon-nucleon potential with an ansatz for the nucleon density in the $\alpha$ cluster. The EDM is calculated as 
\begin{equation}
d_{{}^6\mathrm{Li}} = 0.88(d_n + d_p) + 0.28\,\bar g_1\,e\,\mathrm{fm}+\dots\,,
\end{equation}
where only the nucleon EDM and pion-exchange pieces are kept. The $\bar g_1$ coefficient is $50\%$ larger than that of the deuteron due to the addition $\alpha$-nucleon CP-odd force. More recently, a larger set of light nuclear EDMs, up to ${}^{19}$F, were computed in the no-core shell model using $\chi$EFT CP-conserving and CP-odd one-meson-exchange potentials (\cite{Froese:2021civ}). No systems with large enhancements over the deuteron were identified. 

It is now possible to see if nuclear EDMs are indeed enhanced over nucleon EDMs. Let's take the deuteron EDM as an example and consider the qCEDM as the CP-violating source. From the NDA estimates in Table \ref{tab:NDA}, we expect $(d_n + d_p) \sim e \tilde d_q (F_\pi/\Lambda_\chi)$ while $0.19\, \bar g_1\, e\,\mathrm{fm}   \simeq  e \tilde d_q $. While we used NDA estimates for the LECs a similar conclusion is reached when using QCD sum rules. So indeed, the nuclear force contribution is expected to dominate over the sum of the nucleon EDMs by a factor $\Lambda_\chi/F_\pi$ if the qCEDM is the underlying mechanism. If on the other hand, the qEDM is the dominant source we would expect the deuteron EDM to be approximated by the sum of the nucleon EDMs. The fact that different sources of CP violation leads to different ratios of nuclear-to-nucleon EDMs implies that the source can be unraveled from measurements on different systems (\cite{Dekens:2014jka}).

\section{Diamagnetic electric dipole moments and Schiff's theorem}\label{sec:dia}

While EDMs of light nuclei are interesting from a theoretical point of view, at present the experimental sensitivities are not impressive. Atoms and molecules are much easier to manipulate in the laboratory and resulting EDM limits on these neutral systems are very strong. For example, the limit on the EDM of the diamagnetic atom (with closed electron shells) $d_{{}^{199}\mathrm{Hg}} < 7.4 \cdot 10^{-30}\,e\,\mathrm{cm}$ (\cite{PhysRevLett.116.161601}). While this limit is 2500 times stronger than that on the neutron EDM, a direct comparison is misleading because of electron screening effects that suppress contributions from nuclear CP violation to diamagnetic atomic EDMs. This electron screening, often called Schiff shielding (\cite{Schiff:1963zz}),  can classically be understood from the fact that a neutral bound system composed of charged constituents does not move in presence of an external electric field. That is, the constituents rearrange in such a way that the center of mass feels a vanishing electric field. This theorem can be made exact in quantum mechanics for neutral systems of non-relativistic point particles, see \cite{Engel:1999np} for a pedagogical derivation. 

In diamagnetic systems Schiff screening is (somewhat) avoided by the fact that nuclei are extensive objects and this leads to nonzero atomic EDMs. The important quantity, discussed below, is in this case the nuclear Schiff moment, see \cite{Engel:2025uci} for a recent review. Additional contributions arise from CP-odd electron-nucleus interactions and, for nuclear spin $\geq 1$, from higher CP-odd nuclear moments such as the magnetic quadrupole moments (\cite{Flambaum:1994upo}). While the latter can actually dominate certain systems, the main attention in the field has been on the Schiff moment contributions to the atomic EDM. 

 We can express the Schiff moment of a nucleus $A$ through a combination of three nuclear quantities
\begin{equation}\label{SchiffExpression}
S_A = \frac{d_A}{6}\left(\langle r^2\rangle_{\mathrm{EDM}} -  \langle r^2\rangle_{\mathrm{charge}} \right),
\end{equation}
in terms of the electric dipole radius $\langle r^2\rangle_{\mathrm{EDM}}$, the charge radius $\langle r^2\rangle_{\mathrm{charge}}$, and the nuclear EDM $d_A$. The electric dipole radius is defined as the slope of the electric dipole form factor, see Eq.~\eqref{EDFF}, at $Q^2=0$
\begin{equation}
\langle r^2\rangle_{\mathrm{EDM}} = \frac{6}{d_A} \frac{dF_{\mathrm{EDM}}(Q^2)}{dQ^2}\bigg{|}_{Q^2=0}\,,
\end{equation}
in analogy to the charge radius. Eq.~\eqref{SchiffExpression} shows that the Schiff moment vanishes if the electric dipole radius equals the charge radius. 

Although experimentally not interesting it can be illuminating to take the hydrogen atom as a case study. While hydrogen is not diamagnetic, having a single unpaired electron, it is still useful to see how the Schiff moment affects the atomic EDM. Since the proton charge radius scales as $\langle r^2\rangle_{\mathrm{charge}}\sim 1/\Lambda_\chi^2$ (\cite{Bernard:1995dp}) while the electric dipole radius scales as $d_p/m_\pi^2$ (see Eqs.~\eqref{eq:dnpchiral} and \eqref{nucleonschiff}) we observe that for the proton\footnote{This is true for CP-odd sources that break chiral symmetry such as the $\bar \theta$ term. For the qEDM and Weinberg operator the electric dipole radius scales as $d_p/\Lambda_\chi^2$.} $S_p \sim d_p/m_\pi^2$. The Schiff moment leads to a CP-odd atomic Hamiltonian of the form $H_S \sim  e S_A \vec \sigma \cdot \vec \nabla \delta^{(3}(\vec r)$ arising from a photon exchange between the atomic electron and the proton Schiff moment. The contact nature can be readily understood in momentum space as the Schiff moment involves two derivatives and the resulting $\vec q^{\,2}$ cancels the photon propagator resulting in a contact interaction (\cite{Thomas:1994wi}). The atomic EDM can then be computed in perturbation theory
\begin{equation}
d_{{}^1\mathrm{H}} \sim \sum_{n>1} \frac{\langle 0 |e r |n\rangle \langle n |H_S|0 \rangle}{E_0-E_n}\,,
\end{equation}
where $|0\rangle$ is the hydrogen ground state and $|n \rangle$ excited opposite parity states. A quick calculation for $n=2,\,l=1$ states gives the scaling 
\begin{equation}
d_{{}^1\mathrm{H}} \sim \alpha_{\mathrm{em}}^2 m_e^2 S_A \sim \alpha_{\mathrm{em}}^2 \frac{m_e^2}{m_\pi^2} d_p\,,
\end{equation}
such that the hydrogen EDM is smaller than the proton EDM by a factor $\alpha_{\mathrm{em}}^2 m_e^2/m_\pi^2 \sim 10^{-9}$. Schiff screening is very severe. For the deuteron, the $\chi$EFT calculations with perturbative pions of the charge radius (\cite{Kaplan:1998sz}) and electric dipole form factor (\cite{deVries:2011re}) show that $S_d \sim d_D/\gamma^2$ where $\gamma \simeq m_\pi/3$ is the deuteron binding momentum. The deuterium-to-deuteron EDM is therefore somewhat enhanced over the hydrogen-to-proton EDM. 

The arguments above show that atomic EDMs are roughly suppressed by the square of the size of the nucleus over the size of the atom. Naively already for hydrogen-like atoms, $\alpha_{\mathrm{em}} \rightarrow Z \alpha_{\mathrm{em}}$, and the amount of screening drops with $Z^2$. In addition, electrons become relativistic for larger $Z$ leading to an increased electron probability density at the nucleus, further reducing the screening to roughly the $10^{-3,-4}$ level explaining why experimental searches target high-$Z$ atoms. Much more detailed calculations can be found in (\cite{Dzuba:2002kg, Ginges:2003qt}). We write
\begin{equation}
d_A = \kappa_A S_A \,,
\end{equation}
and give coefficients for several systems \begin{equation}
\kappa_{{}^{129}\mathrm{Xe}} =3.8\cdot10^{-5}\,\mathrm{fm}^{-2}\,,\qquad \kappa_{{}^{199}\mathrm{Hg}} =-2.8\cdot10^{-4}\,\mathrm{fm}^{-2}\,,\qquad \kappa_{{}^{225}\mathrm{Ra}} =-7.7\cdot10^{-4}\,\mathrm{fm}^{-2}\,,
\end{equation}
confirming that Schiff screening diminishes for larger systems a bit faster than $Z^2$. 

The crucial task is then to compute the nuclear Schiff moment $S_A$ in terms of the CP-odd hadronic interactions in Eq.~\eqref{eq:LECs}. While light nuclear EDMs have been computed in terms of CP-odd pion-exchange, the three-body force, and short-range interactions, Schiff moments have focused on the pion-exchange pieces. A rough estimate (for nuclei that are not octupole deformed) in terms of $\bar g_{0,1}$ can be obtained by taking a nuclear-mean field approximation and treating the CP-odd pion-exchange as a short-range contact potential. This leads to an estimate (\cite{Flambaum:1984fb, Engel:2025uci})
\begin{equation}\label{simpleSchiff}
S_A \simeq 0.06\,A^{2/3} \left(\frac{N-Z}{A} \bar g_0 - \bar g_1\right)\,e\,\mathrm{fm}^3\,.
\end{equation}
For heavy nuclei, the prefactor $0.06 A^{2/3} = \mathcal O(1)$, while $(N-Z)/A \simeq 1/5$. For example, for ${}^{199}$Hg we obtain
\begin{equation}
d_{{}^{199}\mathrm{Hg}} \simeq \kappa_{{}^{199}\mathrm{Hg}} \left(-2 \bar g_1 + 0.4 \bar g_0 \right)\,e\,\mathrm{fm}^3\simeq 5\cdot 10^{-4} \left(\bar g_1 - 0.2 \bar g_0\right)\,e\,\mathrm{fm}\,,
\end{equation}
and thus the $\bar g_1$ coefficient is smaller by three orders of magnitude than that of the unscreened deuteron. The simple estimate in Eq.~\eqref{simpleSchiff} is probably an overestimate as it misses the effects beyond valence nucleons. The so-called `core polarization' (\cite{Flambaum:1985gv,Dmitriev:2004fk}) gives rise to additional contributions to the Schiff moment. Explicit calculations in various nuclear methods show that they tend to reduce the coefficients, in particular for $\bar g_1$, sometimes even changing the sign. These cancellations lead to a significantly larger theoretical uncertainty than those affecting EDMs of light nuclei. A compilation of results (\cite{Flambaum:1984fb, Dmitriev:2004fk,Ban:2010ea,Yanase:2020agg})  taken from the review (\cite{Engel:2025uci}) gives
\begin{equation}
d_{{}^{199}\mathrm{Hg}} \simeq \kappa_{{}^{199}\mathrm{Hg}} \left[ (0.6\pm 0.6) \bar g_0 +  (0.4 \pm 0.8)\bar g_1 \right]\,e\,\mathrm{fm}^3\,, 
\end{equation}
where the uncertainty is arising from the spread in various calculations. The fact that the range includes very small or opposite sign coefficients is worrisome. Furthermore, taking the spread of calculations is hardly a proper estimate of the theoretical uncertainty. 

The Schiff moment is also affected by the other LECs in Eq.~\eqref{eq:LECs}. While the short-range forces have not been systematically studied there are contributions from the nucleon EDMs (\cite{Dmitriev:2003sc}) and the semi-leptonic tensor operator $C_T$ (\cite{Latha:2009nq}), modifying the EDM expression into
\begin{equation}
d_{{}^{199}\mathrm{Hg}} \simeq \kappa_{{}^{199}\mathrm{Hg}} \left\{ \left[(1.9 \pm 0.1)d_n + (0.20\pm0.06)d_p\right]\,\mathrm{fm}^{-1} + (0.6\pm 0.6) \bar g_0 +  (0.4 \pm 0.8)\bar g_1 \right\}\,e\,\mathrm{fm}^3 + (1.2\pm 0.2) (C_T^{(0)} - C_T^{(1)}) \cdot 10^{-7}\,e\,\mathrm{fm}\,. 
\end{equation}
The indirect limit on the proton EDM is obtained by assuming the other contributions to vanish giving $d_p \lesssim 10^{-25}$ e cm. However, it is difficult to envision a BSM scenario where $d_p$ is actually the dominant contribution considering the larger neutron EDM coefficient. In addition, the relative contributions from the nucleon EDMs and CP-odd pion exchange is similar to that of light nuclei, and we can again expect that for sources such as the qCEDM and the FQLR, the $\bar g_{0,1}$ contributions are dominating. 

The simple estimate in Eq.~\eqref{simpleSchiff} does not work for octupole deformed nuclei. In particular, ${}^{225}$Ra has a close-lying excited nuclear state with a splitting of order tens of keV instead of the typical MeV nuclear splittings (\cite{Spevak:1995zem, Auerbach:1996zd}). \cite{Dobaczewski:2018nim} showed that there is a correlation between the ${}^{225}$Ra Schiff moment and the ${}^{224,226}$Ra octupole moments. The latter are measured guiding the Schiff moments calculations leading to
\begin{equation}\label{eq:dRA}
d_{{}^{225}\mathrm{Ra}} \simeq \kappa_{{}^{225}\mathrm{Ra}} \left[ (2.5\pm 7.5) \bar g_0 -  (64 \pm 37)\bar g_1 \right]\,e\,\mathrm{fm}^3\,,
\end{equation}
showing much larger coefficients. Despite this enhancement atomic ${}^{225}$Ra EDM experiments (\cite{Bishof:2016uqx}) are not sufficiently precise to compete with ${}^{199}$Hg, but experiments with molecules containing octupole deformed nuclei such as RaF are very promising (\cite{Jadbabaie:2026njt}). 
Recently the first beyond-mean-field calculation (\cite{Zhou:2025jfi}) of the ${}^{225}$Ra (and other isotopes) Schiff moment was performed finding somewhat smaller coefficients than Eq.~\eqref{eq:dRA}. The calculation also uncovered a correlation between the Schiff moment contributions of nuclear intermediate states and their electric dipole transition strengths to the ground state. This connection allows experimental measurements of these transitions to constrain the nuclear models used to calculate the Schiff moments.

In the future it might become possible to perform first-principle nuclear Schiff moment calculations using $\chi$EFT in similar spirit as the light-nuclear EDM calculations. Recently, the Schiff moment of ${}^{19}$F was computed with $\chi$EFT wave functions using the no-core shell model (\cite{Ng:2025hgx})
\begin{equation}
S_{{}^{19}\mathrm{F}}= \left[ -(0.38\pm0.19)\,\bar g_0   - (0.31\pm0.16)\,\bar g_1\right]\,e\,\mathrm{fm}^3\,,
\end{equation}
with $50\%$ uncertainties on the coefficients. The $S_{{}^{19}\mathrm{F}}$ can be constrained using the polar molecular measurements on TlF, YbF, and HfF$^+$, but, because, the electron density peaks at the heavy nucleus contained in the molecule, the resulting limits are not yet competitive. Nevertheless, the results show the power of ab initio computations and pave the way towards calculations on larger systems. One such calculation was recently performed (\cite{Belley:2026bde}) using the in-medium similarity renormalization group (\cite{Hergert:2015awm}) where Schiff moments of several light-to-medium heavy nuclei were computed in addition to a calculation of 
\begin{equation}
S_{{}^{129}\mathrm{Xe}}= \left[ -0.27\,\bar g_0   - 0.13\,\bar g_1\right]\,e\,\mathrm{fm}^3\,.
\end{equation}
The current limit is $d_{{}^{129}\mathrm{Xe}} < 1.5 \cdot 10^{-27}$ e cm (\cite{PhysRevA.100.022505}) and thus the sensitivity is not competitive yet with the ${}^{199}$Hg limits. The xenon calculations are based on a single nuclear interaction and no theoretical uncertainty was given. Cross-checks with other interactions or methods are needed. 

The development of first-principle nuclear calculations of symmetry-breaking moments is a very promising direction, but significant work reamins. Calculations have focused solely on the CP-odd one-pion-exchange contributions, but already in light nuclei the two-pion-exchange diagrams (\cite{Maekawa:2011vs}) are sizable (\cite{Gnech:2019dod}). The effects from the CP-odd three-pion vertex $\bar \Delta$ have not been included in any Schiff moment calculation and the role of CP-odd short-range forces is not systematically investigated. In particular in light of the renormalization issues discussed in Sect.~\ref{renorm} the last issue is very pressing. Another open issue is the systematic calculation of nuclear magnetic quadrupole moments (MQM). The $\chi$EFT calculation of the deuteron MQM (\cite{Liu:2012tra}) shows that additional CP-odd hadronic interactions can play a role. Since in several system the nuclear MQM is expected to dominate atomic EDMs (\cite{Flambaum:1994upo,Flambaum:2014jta}), ab initio calculations of MQMs would be very interesting.

\section{Paramagnetic electric dipole moments}

In paramagnetic systems there is an unpaired electron leading to net electron spin. While Schiff's theorem argues that the electron EDM contribution to the entire neutral system is screened, this can be avoided if relativity is taken into account (\cite{Sandars:1965gzg, SANDARS1966290}). In large atoms or molecules, even valence electrons are relativistic and Schiff's screening can be completely avoided or even overturned: the EDM of a large atom can be larger than that of the electron. Explicit calculations of the enhancement factors are discussed in \cite{Ginges:2003qt}. In heavy systems the largest effect arises from the electron EDM interacting with the nuclear charge, leading to a small admixture of opposite parity electron states. Explicit calculations show that this effect leads to a ratio of atomic-to-electron EDM
\begin{equation}
\frac{d_A}{d_e} \equiv K \sim \alpha^2 Z^3 f(Z \alpha) \,,
\end{equation}
where $f$ is a monotonically increasing function. As such, the enhancement factor grows 'faster than $Z^3$'. For example, for ${}^{133}$Cs explicit calculations give $K(\mathrm{Cs})= 114$ (\cite{Hartley:1990qn}) and for ${}^{205}$Tl $K(\mathrm{Tl})= -573$ (\cite{Porsev:2012zx}). Similarly, contributions from the CP-violating scalar nucleon-electron coupling, $C_S$, in Eq.~\eqref{slh} are enhanced and are tightly constrained by paramagnetic EDM limits.  

Atomic EDMs experiments historically set the strongest electron EDM limits but have now been superseded by molecular measurements. In typical experimental electric fields, atoms are only very weakly polarized because the applied electric field is tiny compared to the internal atomic field that binds the valence electron. In a polar molecule however, there is a near-degenerate opposite parity state with an energy splitting small enough that even the weak laboratory field is sufficient to fully mix them and to completely orient the internuclear axis. This leads to an extra enhancement of $M_{\mathrm{mol}}/m_e$ (\cite{Sushkov:1978yj}), an astonishing  factor of  $\mathcal O(5\cdot 10^5)$ for a heavy molecule like ThO,   where the molecular mass enters because the doublet splitting is set by the rotational energy scale. Experiments on polar molecules such as YbF (\cite{Hudson:2011zz}), ThO (\cite{ACME:2018yjb}), and HfF$^+$ (\cite{Roussy:2022cmp}) have increased rapidly in the last 15 years and are now by far the best probe of (semi-)leptonic CP violation. 

For atoms the induced precession frequency of the atomic spin is linearly dependent on the applied external electric field. For polar molecules, the valence electron feels the internal electric field which is a molecule-dependent quantity that saturates at weak fields and is largely independent of the external field once the molecule is fully polarized. As such, experiments do not report a limit on molecular EDMs but on the observed frequency shift. This frequency shift can be expressed in contributions from the electron EDM, $d_e$, and CP-violating electron-nucleon interactions. For example
\begin{eqnarray}
\omega_{\text{HfF}} &=&(34.9\pm1.4)(\mathrm{mrad}/\mathrm{s})\left(\frac{d_e}{10^{-27}\,e\,\mathrm{cm}}\right)+(32.0\pm1.3)(\mathrm{mrad}/\mathrm{s})\left(\frac{C_S }{10^{-7}}\right)\,,
\end{eqnarray}
where the coefficients in front of $d_e$ and $C_S$ are molecular matrix elements obtained with relativistic many-body calculations (\cite{Skripnikov:2017cnj,Fleig:2018bsf, Haase_2021}). Assuming $C_S=0$ and using the limit on $\omega_{\mathrm{HfF}}$ (\cite{Roussy:2022cmp}) then leads to $d_e < 4.1 \cdot 10^{-30}$ e cm. Since limits on EDMs are easier to interpret than limits on frequencies, \cite{Pospelov:2013sca} suggested to interpret paramagnetic EDM measurements in terms of the `equivalent' electron EDM defined as 
\begin{equation}\label{eq:deequiv}
d_e^{\mathrm{equiv}} = d_e + r_A C_S\,,
\end{equation}
where $r_A$ is a molecule dependent ratio of molecular matrix elements, for example, 
\begin{eqnarray}
r_\text{BaF}&=&\,4.5\cdot 10^{-21} \text{ e cm}\,,\qquad
    r_\text{ThO}=\,1.5 \cdot 10^{-20} \text{ e cm}\,,\qquad
        r_{\text{HfF}^+}=9.2\cdot 10^{-21} \text{ e cm}\,.
\end{eqnarray}
The fact that these ratios are not precisely the same implies that measurements on several systems can unravel the electron EDM from $C_S$. 

\subsection{Using paramagnetic EDMs to probe hadronic CP violation }\label{sec:paratodia}

The discussion so far has mainly focused on the electron EDM and the CP-odd electron-nucleon coupling $C_S$ as the quantities directly constrained by paramagnetic EDM experiments. While the electron EDM can be computed directly in  the SM or in beyond-the-SM scenarios, $C_S$ is an effective hadronic interaction that  
 must  be expressed in terms of more fundamental source of CP violation. The most obvious source are the four-fermion electron-quark interactions in Eq.~\eqref{Lsl}. In this way, stringent limits on CP violation in, for example, leptoquark models can be derived  (\cite{Dekens:2018bci}).

 \begin{figure}[t!]
    \centering
    \includegraphics[width=0.7\textwidth]{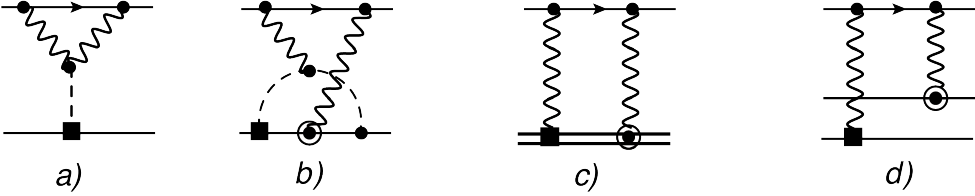}
    \caption{Several diagrams contributing to a CP-violating electron-nuclear interaction. The top solid arrowed line is the electron and the other solide lines are nucleons. The double line in diagram c) denotes the nucleus as a whole. Wavy and dashed lines are photons and pions. The black square is a CP-odd interaction while black circels denote CP-even interactions. The circled circle in diagram b), c), d) denotes the nucleon or nuclear magnetic moment. }
    \label{fig:SL}
\end{figure}

 More intricate contributions to $C_S$ have been the focus of more recent studies. The key observation (\cite{Flambaum:2019ejc,Mulder:2025esr,Dekens:2025skl}), is that paramagnetic experiments are also sensitive to hadronic CP violation and that this sensitivity is complementary to and competitive with dedicated hadronic EDM searches. The main physical mechanism is the following. As discussed in Sect.~\ref{sec:chiralEFT}, sources of hadronic CP violation that violate chiral symmetry, such as the $\bar\theta$ term, qCEDMs, or the FQLR, generate CP-odd pion-nucleon couplings $\bar g_{0,1}$. The CP-odd pion-nucleon couplings generate a virtual neutral pion in the system which then couples to two photons through the chiral anomaly and the photons, in turn, couple to an electron (see Fig.~\ref{fig:SL}a)). The resulting effective interaction is precisely the scalar-pseudoscalar electron-nucleon coupling $C_S$ of Eq.~\eqref{slh}. 

For $\bar g_1$ this effect is coherent over all nucleons in the nucleus and thus picks up an overal factor $A$ while for $\bar g_0$ a relative $(Z-N)/A$ suppression appears. This implies that for the $\bar \theta$ term, for which $\bar g_0$ is the only leading-order CP-violating interaction, formally next-to-leading order interactions are relevant. These involve isospin-breaking corrections through $\bar g_1$, strangeness corrections through $\eta$ exchange, and two-loop pion-photon loops (Fig.~\ref{fig:SL}b)). Combined (\cite{Mulder:2025esr}) gives the following result for the nucleus-averaged coupling
\begin{equation}\label{Cspion}
C_S = \frac{\sqrt{2}}{G_F} \frac{\alpha^2}{4 \pi^2} \frac{ m_e}{ F_\pi m_\pi^2} \bigg\{ \left[\frac{Z-N}{Z+N} \bar{g}_0 + \bar g_1 \right]\mathcal{B}^\pi + \frac{F_\pi}{\sqrt{3} F_\eta}\frac{ m^2_\pi}{ m_\eta^2} \bar{g}_{0\eta} \mathcal{B}^\eta + \bar{g}_0\frac{\pi g_A m_\pi }{m_N} \frac{ Z \mu_n-N \mu_p}{(Z+N)} \left[\log{\left(\frac{m_\pi}{m_e}\right)}+1.77 \right] \bigg\}\,,
\end{equation}
 where $B^{\pi}\simeq B^{\eta} \simeq 45$ are renormalized one-loop diagrams, and $\mu_n = -1.91$, $\mu_p = 2.79$ are the nucleon magnetic moments in units of the nuclear magneton. The term in brackets proportional to the magnetic moments is the result of a numerical evaluation of a two-loop integral. 
 Plugging in values of the $\bar g_{0,1}$ and $\bar g_{0\eta}$ (taken from \cite{deVries:2015una}) for $\bar \theta$ and using typical $N,Z$ values gives 
\begin{equation}\label{Csthetapion}
C_S(\bar \theta)= (1.63 \pm 0.45) \cdot 10^{-2}\,\bar \theta \,,
\end{equation}
or, equivalently, for HfF${}^+$, 
\begin{equation}
    d_e^\text{equiv}(\bar\theta) = (1.50\pm 0.42) \cdot 10^{-22} \,\bar{\theta} 
\text{ e cm}\,,
\end{equation}
and thus the paramagnetic bound gives $\bar \theta <1.5\cdot 10^{-8}$ roughly two orders of magnitude weaker than the traditional limit obtained from the neutron EDM. This bound does not include contributions from nucleon EDMs discussed below. In similar fashion, bounds can be set on other CP-violating hadronic sources. For the qCEDM and FQLR, the $\bar g_1$ contribution is dominant and the other terms in Eq.~\eqref{Cspion} can be neglected.   

Additional contributions to paramagnetic EDMs can arise from a combination of nucleon electric and magnetic dipole moments (\cite{Flambaum:2019ejc}, \cite{Dekens:2025skl}). These contributions, shown in Figs.~\ref{fig:SL}c) and \ref{fig:SL}d) are particularly relevant for hadronic sources, such as the qEDM or the Weinberg operator, for which the CP-violating pion-nucleon couplings are suppressed. For diagram \ref{fig:SL}c) it is no longer valid to talk about CP-odd interactions between electrons and individual nucleons, but rather between electrons and the nucleus as a whole. I denote this coupling by $\bar C_S$. In the first diagram, the exchanged photons have virtuality of the order of nuclear excitation energies $\Delta_n$ around a few MeV. Under the assumption that $\Delta_n \gg m_e$, the expression for $\bar C_S$ becomes compact 
\begin{align}\label{CSPeffTOT}
\bar C_S =-\frac{\sqrt{2}}{G_F}\frac{4\alpha^2 m_e}{m_N}\sum_n \frac{A_n}{\Delta_n}\left(3\ln \frac{m_e^2}{4\Delta_n^2}-1\right)\,,\qquad A_n =- \frac{\langle h_i | D \vec \sigma | n \rangle \cdot \langle n |  \mu \vec \sigma |h_i \rangle}{12}\,.
\end{align}
Here $A_n$ is a nuclear matrix elements between the ground state $|h_i\rangle$ and  $1^+$ nuclear excited states $|n\rangle$ involving the EDM ($D$) and magnetic dipole ($\mu$) operator. The calculation of $A_n$ is difficult as it requires a sum over many states. An explicit nuclear shell model calculation was performed for BaF (\cite{Dekens:2025skl}) for which a first limit was recently reported (\cite{Boeschoten:2026inq}) leading to
\begin{equation}\label{usoft}
\bar C_S(\mathrm{BaF}) = (67 \pm 28) \frac{d_p}{e\,\mathrm{fm}}\,.
\end{equation}
Most of the contributions arises from $1^+$ excited states with energies around 4 to 5 MeV and higher states contribute little. The absence of a neutron EDM contribution is specific to ${}^{138}$Ba which has a magic neutron number. It would be interesting to perform similar computations for the larger ThO and HfF${}^+$ systems.  

Diagram \ref{fig:SL}d) involves two separate nucleons in a nucleus. The resulting value of $\bar C_S$, the effective CP-odd electron-nucleus coupling, involves the calculations of a ground-state to ground-state two-nucleon matrix element 
\begin{align}\label{CSPeffTOT}
\bar C_S =-\frac{\sqrt{2}}{G_F}\langle h_i| V|h_i\rangle\,,\qquad V =\frac{4e^4 m_e}{9m_N } \sum_{i\neq j}\frac{ \mu^{(i)} D^{(j)} }{|\vec q|^4}\left[ \vec \sigma^{(i)}\cdot  \vec \sigma^{(j)}-\frac{1}{4}S^{(ij)}\right]\,,
\end{align}
where $\vec q$ is the momenum transfer and $S^{(ij)} =\vec \sigma^{(i)}\cdot \vec \sigma^{(j)} -3 (\vec q\cdot \vec \sigma^{(i)})\, (\vec q \cdot \vec \sigma^{(j)})/|\vec q|^2$ is a tensor operator. Explicit nuclear shell model calculations for a range of nuclei show that the effect is coherent and scales with the number of neutrons and protons. For BaF in particular 
\begin{equation}\label{potBa}
\bar C_S(\mathrm{BaF}) = \left[(-433 \pm 5) \, d_p + (387\pm 0.4)\, d_n \right]\,(e\, \mathrm{fm})^{-1}\,,
\end{equation}
several times larger than Eq.~\eqref{usoft}. The coherence can be used to compute the result for a general polar molecular containing a heavy nucleus $A$ with $Z$ protons and $N$ neutrons
 \begin{equation}
    {d}^{\rm equiv}_e=\frac{\sqrt{2}e^4m_e}{18\pi G_Fm_N} \frac{r_A}{A} (-9.7) \left[  \, Z \, \mu_p\, d_p+N \,\mu_n \,d_n \right] \frac{\mathrm{fm}}{e} \,,
    \label{eq:masterformula}
\end{equation}
 where the $-9.7$ is a numerical coefficient computed with the shell model with an expected $25\%$ uncertainty mainly from higher-order chiral corrections. This formula makes it possible to constrain nucleon EDMs from the most precise HfF${}^+$ measurement. Because $Z \mu_p \simeq - N \mu_n $ for heavy nuclei, the paramagnetic molecules mainly constrains the isovector combination 
\begin{equation}
|d_n - d_p| < 1.6 \cdot 10^{-23}\,e\,\mathrm{cm}\,. 
\end{equation}
The direct neutron EDM limit is stronger by three orders of magnitude, while the inferred proton EDM limit from the ${}^{199}$Hg measurement is a hundred times stronger. 

The above discussion shows that paramagnetic EDM measurements are becoming 'diamagnetic' in the sense that they can constrain CP-violating sources in the quark-gluon sector through distinct and calculable mechanisms. The interpretation of future measurements in terms of all these contributions simultaneously will require combining results from several molecular systems with different $r_A$ ratios and will demand continued progress in both nuclear structure calculations and chiral EFT matching. While in absolute limits there is  still a gap of two-to-three orders of magnitude in sensitivity, the paramagnetic EDM limits have made much faster progress and a further reduction of the gap is definitely possible. Perhaps more importantly, the experiments are complementary as the ratio of paramagnetic-to-diamagnetic EDMs can be used to identify the underlying CP-violating mechanism. This will be discussed in the next section. 

\section{Unraveling the mechanism of CP violation with the EDM portfolio}

While squarely in the category of problems one hopes to face, a measurement of a nonzero EDM immediately raises the question of which source is responsible. Assuming the discovery is made in the foreseeable future, we can rule out the CKM mechanism (\cite{Pospelov:2013sca}), leaving either the $\bar\theta$ term or a BSM source of CP violation as the explanation.
Identifying the $\bar\theta$ term as the source would have far-reaching consequences for the strong CP problem. 

Solutions to the strong CP problem fall into two broad classes~(\cite{Craig:2022eqo}): UV solutions, where CP or P is an exact symmetry that is spontaneously broken at high energies~(\cite{Nelson:1983zb,Barr:1984qx,Babu:1989rb}), and IR solutions such as the Peccei-Quinn mechanism~(\cite{Peccei:1977hh}), where $\bar\theta$ is relaxed to zero dynamically at low energies.
Generically, UV solutions do not predict new sources of hadronic CP violation beyond $\bar\theta$, because the symmetry protecting $\bar\theta$ also suppresses other CP-odd dimension-six operators below observable levels. To be clear, this is a naturalness argument which, as history shows,  have not always been reliable. Nevertheless, from an EFT perspective it is difficult to understand why $\bar\theta$ remains small in the IR if other hadronic CP-violating sources are present. A pattern of EDMs consistent with a BSM hadronic source but inconsistent with a $\bar\theta$-dominated scenario would therefore point toward an IR solution, such as the Peccei--Quinn mechanism, as the explanation for the smallness of $\bar\theta$~(\cite{deVries:2021sxz,Choi:2023bou,Choi:2026tun}).
On the other hand, if the source of CP violation is not the $\bar\theta$ term, identifying which BSM operator is responsible becomes equally important as it would inform us about the possible UV completions, guide complementary searches at colliders and low-energy precision experiments, and shed light on viable mechanisms of electroweak baryogenesis.

If nonzero EDMs are measured, the first question is whether the source is (semi-)leptonic or hadronic. This can be determined through the ratio of paramagnetic EDMs over neutron/diamagnetic EDMs. If the paramagnetic EDMs are relatively large, the source is most likely the electron EDM, $d_e$ or the CP-odd electron-nucleon interactions $C_S$. By comparing  the ratio of several paramagnetic systems with different $r_A$ coefficients, see Eq.~\eqref{eq:deequiv}, it should be possible to determine the origin (\cite{Chupp:2017rkp,Fleig:2018bsf}).

\begin{figure}[t!]
    \centering
    \includegraphics[width=0.45\textwidth]{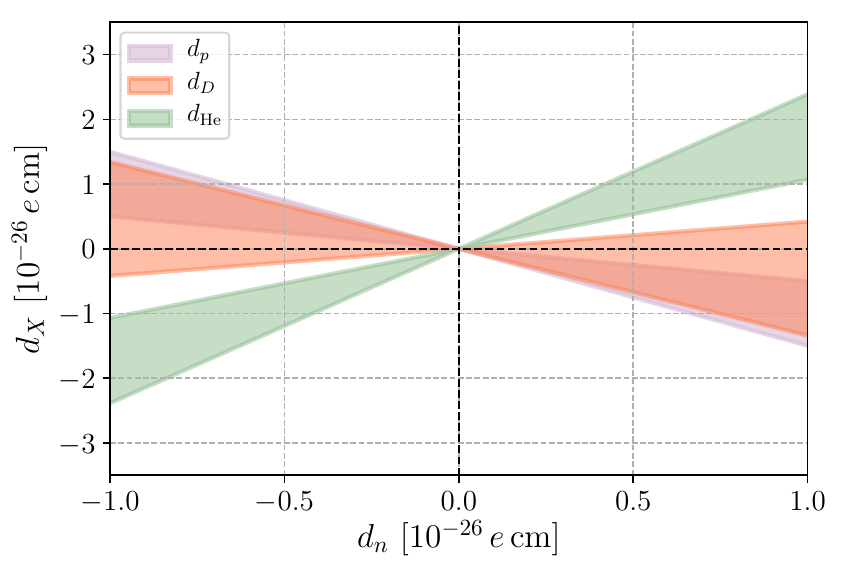}
    \includegraphics[width=0.45\textwidth]{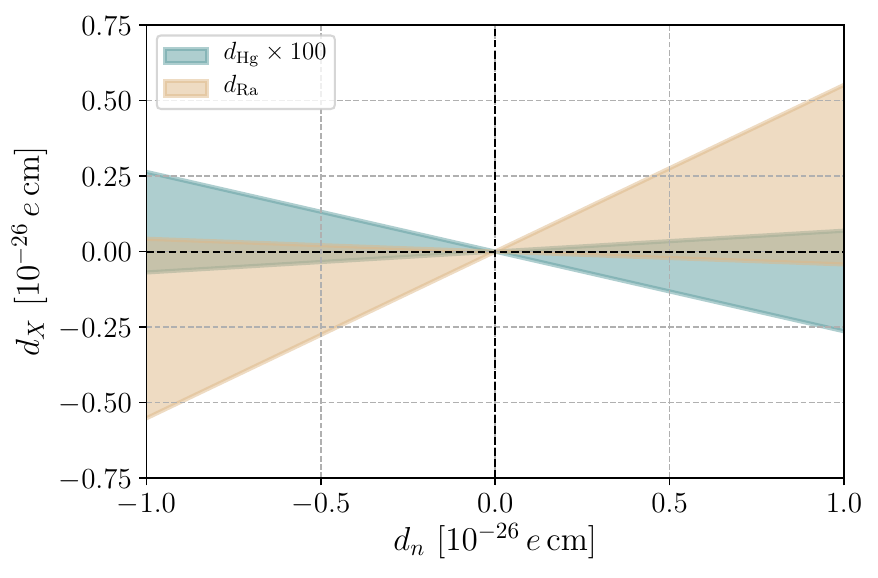}
    \caption{Left panel: Correlation between various EDMs (proton and light nuclei) and the neutron EDM in case the $\bar \theta$ term is the underlying CP-violating mechanism. The bands indicate the theoretical uncertainty. EDM measurements that would fall outside the shared areas, points towards a BSM CP-violating source and, indirectly, towards an IR solutions of the strong CP problem. Right panel: the same but now for EDMs of diamagnetic EDMs versus the neutron EDM. Here the nuclear uncertainty typically leads to broader uncertainty bands. Figure adapted from~\cite{deVries:2018mgf}.}
    \label{fig:ratio}
\end{figure}

If the source is hadronic, the first question should be whether the pattern of EDMs is consistent with the $\bar \theta$ term. All calculations indicate the nucleon EDMs induced by $\bar \theta$ are mainly isovector and thus $d_n$ and $d_p$ appear with opposite sign. In addition, this implies that the deuteron EDM is relatively small because both $d_n+d_p$ and $\bar g_1$ are small for $\bar \theta$, while ${}^3$He can be expected to be larger. This pattern of nucleon and light nuclear EDMs is shown in the left panel of Fig.~\ref{fig:ratio}. However experiments probing EDMs of light nuclei are currently not competitive. The limit on the ${}^{199}$Hg EDM is very strong but the large nuclear uncertainties of the $\bar g_0$ coefficient makes a determination more difficult. The same is true for ${}^{255}$Ra, see the right panel of Fig.~\ref{fig:ratio}. Interestingly, paramagnetic systems, through Eq.~\eqref{Csthetapion} and Eq.~\eqref{eq:masterformula}, can also be used to identify the presence of $\bar \theta$ as it predicts, for example, the $\omega_{\mathrm{HfF}}/d_n$ ratio (\cite{Dekens:2025skl}).

If the source is not $\bar \theta$, there are several options. The qCEDM and FQLR predict that nuclear and diamagnetic EDMs are dominated by the CP-odd nuclear forces mainly through a large $\bar g_1$ coefficient (and possibly $\bar \Delta$ although the role of three-nucleon CP-odd forces is unclear). This means that these sources tend to predict larger $d_D/d_n$ and $d_{\mathrm{Ra}}/d_n$ ratios than the $\bar \theta$ term. The qEDM and Weinberg operator predict that nuclear and diamagnetic EDMs are mainly functions of the nucleon EDMs and no large ratios are expected. 

The above predictions are subject to substantial hadronic and nuclear uncertainties. For sources beyond $\bar\theta$ and qEDMs, the CP-odd LECs are currently known only at the order-of-magnitude level, and the role of short-range CP-odd nuclear forces, three-body interactions, and subleading pion-exchange corrections remains poorly understood beyond light nuclei. Reducing these uncertainties,  through lattice QCD determinations of the hadronic LECs and ab initio nuclear calculations for heavier systems,  is essential if a future EDM signal is to be unambiguously traced back to its origin.

\section{Conclusion}

This review has presented the theory of electric dipole moments from a low-energy perspective, 
tracing the chain of connections that links CP-violating interactions at the quark and 
gluon level to observable EDMs of nucleons, nuclei, atoms, and molecules. Starting from 
a general CP-odd effective Lagrangian at the hadronic matching scale, comprising the 
$\bar\theta$ term, quark EDMs and chromo-EDMs, the Weinberg operator, and CP-odd 
four-fermion interactions, chiral perturbation theory organizes the 
nonperturbative QCD dynamics into a small set of hadronic LECs. The nuclear EDMs and Schiff moments of light and heavy nuclei were then 
discussed within chiral EFT, emphasizing the power counting that identifies the leading 
CP-odd pion-exchange mechanisms and the role of short-range CP-odd forces whose status is still unclear. At the atomic and 
molecular level, we covered both diamagnetic systems, where sensitivity to 
hadronic CP violation is mediated by nuclear Schiff moments and related CP-odd 
moments, and paramagnetic systems, where the dominant sensitivity is to the electron 
EDM and the scalar electron-nucleon coupling $C_S$. We discussed how hadronic CP violation can
enter paramagnetic systems through a recently identified and still largely unexplored set of mechanisms. 
The complementarity of the full EDM portfolio, encompassing nucleons, light nuclei, 
diamagnetic atoms, and polar molecules, in disentangling the underlying source of 
CP violation was discussed in the final section.

While not covered in this review, the experimental prospects for EDM searches are excellent. Sensitivity to paramagnetic EDMs has already improved by more than two orders of magnitude in the past 15 years. Looking ahead, technological advances across all categories of EDM experiments, from improved neutron EDM measurements, to novel molecular cooling and trapping techniques, to using radioactive species, to storage ring experiments, promise to further push the sensitivity in upcoming years.

The theory of EDMs faces challenges and opportunities at every layer 
of the hierarchy. At the hadronic level, the determination of the CP-odd LECs for 
sources beyond the $\bar\theta$ term remains a pressing 
open problem. The gradient-flow approach to lattice QCD, for which the perturbative 
one-loop matching of all relevant operators has recently been completed, offers a 
realistic path toward first-principles determinations of these LECs. At the nuclear level, the extension of ab initio many-body methods to heavier 
octupole-deformed nuclei such as ${}^{225}$Ra, and the systematic inclusion of the complete set of CP-odd forces, including short-range and three-body interactions, in nuclear 
calculations of Schiff and magnetic quadrupole moments, is essential to exploit the current and next generation of diamagnetic 
experiments. At the molecular level, the connection between hadronic CP violation and 
paramagnetic observables is only beginning to be mapped out.  The view from below, from the mud of hadronic, nuclear, and molecular physics, will become increasingly important as experimental sensitivity 
improves: a future EDM signal will only reveal the underlying mechanism of CP violation 
if the theoretical chain from elementary particles to observables is quantitatively under control at 
every step.

\section*{Acknowledgements}
I thank Heleen Mulder and Lemonia Gialidi for comments on the manuscript and help with the figures. I am grateful to Emanuele Mereghetti, Wouter Dekens, and Vincenzo Cirigliano for discussions on some of the topics presented here. I thank Bira van Kolck and Rob Timmermans for introducing me to this field all those years ago, and the EDM community at large for providing a rich and collegial research environment.  
JdV is supported by the ERC COG grant CRUNS, 101230525, and  by Dutch Research Council (NWO) in the form of a VIDI grant.

\bibliographystyle{Harvard}
\bibliography{reference}

\end{document}